\documentclass[aps,pra,twocolumn,superscriptaddress,footinbib]{revtex4-1}

\makeatletter
\def\@bibdataout@aps{%
 \immediate\write\@bibdataout{%
  @CONTROL{%
   apsrev41Control,author="08",editor="1",pages="0",title="0",year="1",eprint="1"%
  }%
 }%
 \if@filesw
  \immediate\write\@auxout{\string\citation{apsrev41Control}}%
 \fi
}%
\makeatother 

\usepackage[normalem]{ulem}

\usepackage{mathpazo}
\usepackage{mathtools} 
\usepackage{amsmath}
\usepackage{amsfonts, amssymb, amsthm, bbm, braket, accents}
\usepackage{graphicx}   
\usepackage{subfigure}  
\usepackage{amsbsy} 
\usepackage[usenames,dvipsnames]{color}
\usepackage[bold]{hhtensor} 
\usepackage{natbib}
\usepackage{algpseudocode}

\usepackage{multirow}

\usepackage{soul} 

\usepackage{ifthen}
\newboolean{FinalVersion}
\setboolean{FinalVersion}{false}
\newboolean{JournalVersion}
\setboolean{JournalVersion}{false}

\ifthenelse{\boolean{JournalVersion}}{
  
}{
    \usepackage{mathpazo}
}

\ifthenelse{\boolean{FinalVersion}}{

}{
    \newcommand{\todo}[1]{{\color{Plum} #1}}
    \newcommand{\TODO}{\todo{TODO}}

}

    \usepackage{textcomp}
    \usepackage{graphicx} 
    \usepackage{adjustbox} 
    \usepackage{color} 
    \usepackage{enumerate} 
    \usepackage{amsmath} 
    \usepackage{amssymb} 
    \usepackage{eurosym} 
    \usepackage[mathletters]{ucs} 
    \usepackage[utf8x]{inputenc} 
    \usepackage{fancyvrb} 
    \usepackage{grffile} 
    \usepackage{longtable} 
    \usepackage{booktabs}  

    \definecolor{orange}{cmyk}{0,0.4,0.8,0.2}
    \definecolor{darkorange}{rgb}{.71,0.21,0.01}
    \definecolor{darkgreen}{rgb}{.12,.54,.11}
    \definecolor{myteal}{rgb}{.26, .44, .56}
    \definecolor{gray}{gray}{0.45}
    \definecolor{lightgray}{gray}{.95}
    \definecolor{mediumgray}{gray}{.8}
    \definecolor{inputbackground}{rgb}{.95, .95, .85}
    \definecolor{outputbackground}{rgb}{.95, .95, .95}
    \definecolor{traceback}{rgb}{1, .95, .95}
    \definecolor{red}{rgb}{.6,0,0}
    \definecolor{green}{rgb}{0,.65,0}
    \definecolor{brown}{rgb}{0.6,0.6,0}
    \definecolor{blue}{rgb}{0,.145,.698}
    \definecolor{purple}{rgb}{.698,.145,.698}
    \definecolor{cyan}{rgb}{0,.698,.698}
    \definecolor{lightgray}{gray}{0.5}
    
    \definecolor{darkgray}{gray}{0.25}
    \definecolor{lightred}{rgb}{1.0,0.39,0.28}
    \definecolor{lightgreen}{rgb}{0.48,0.99,0.0}
    \definecolor{lightblue}{rgb}{0.53,0.81,0.92}
    \definecolor{lightpurple}{rgb}{0.87,0.63,0.87}
    \definecolor{lightcyan}{rgb}{0.5,1.0,0.83}
    
    \usepackage{inconsolata}
    \DefineVerbatimEnvironment{Highlighting}{Verbatim}{commandchars=\\\{\},fontfamily=zi4}

    \title{QInfer Tomography Tutorial}

\makeatletter
\def\PY@reset{\let\PY@it=\relax \let\PY@bf=\relax%
    \let\PY@ul=\relax \let\PY@tc=\relax%
    \let\PY@bc=\relax \let\PY@ff=\relax}
\def\PY@tok#1{\csname PY@tok@#1\endcsname}
\def\PY@toks#1+{\ifx\relax#1\empty\else%
    \PY@tok{#1}\expandafter\PY@toks\fi}
\def\PY@do#1{\PY@bc{\PY@tc{\PY@ul{%
    \PY@it{\PY@bf{\PY@ff{#1}}}}}}}
\def\PY#1#2{\PY@reset\PY@toks#1+\relax+\PY@do{#2}}

\expandafter\def\csname PY@tok@gd\endcsname{\def\PY@tc##1{\textcolor[rgb]{0.63,0.00,0.00}{##1}}}
\expandafter\def\csname PY@tok@gu\endcsname{\let\PY@bf=\textbf\def\PY@tc##1{\textcolor[rgb]{0.50,0.00,0.50}{##1}}}
\expandafter\def\csname PY@tok@gt\endcsname{\def\PY@tc##1{\textcolor[rgb]{0.00,0.27,0.87}{##1}}}
\expandafter\def\csname PY@tok@gs\endcsname{\let\PY@bf=\textbf}
\expandafter\def\csname PY@tok@gr\endcsname{\def\PY@tc##1{\textcolor[rgb]{1.00,0.00,0.00}{##1}}}
\expandafter\def\csname PY@tok@cm\endcsname{\let\PY@it=\textit\def\PY@tc##1{\textcolor[rgb]{0.25,0.50,0.50}{##1}}}
\expandafter\def\csname PY@tok@vg\endcsname{\def\PY@tc##1{\textcolor[rgb]{0.10,0.09,0.49}{##1}}}
\expandafter\def\csname PY@tok@m\endcsname{\def\PY@tc##1{\textcolor[rgb]{0.40,0.40,0.40}{##1}}}
\expandafter\def\csname PY@tok@mh\endcsname{\def\PY@tc##1{\textcolor[rgb]{0.40,0.40,0.40}{##1}}}
\expandafter\def\csname PY@tok@go\endcsname{\def\PY@tc##1{\textcolor[rgb]{0.53,0.53,0.53}{##1}}}
\expandafter\def\csname PY@tok@ge\endcsname{\let\PY@it=\textit}
\expandafter\def\csname PY@tok@vc\endcsname{\def\PY@tc##1{\textcolor[rgb]{0.10,0.09,0.49}{##1}}}
\expandafter\def\csname PY@tok@il\endcsname{\def\PY@tc##1{\textcolor[rgb]{0.40,0.40,0.40}{##1}}}
\expandafter\def\csname PY@tok@cs\endcsname{\let\PY@it=\textit\def\PY@tc##1{\textcolor[rgb]{0.25,0.50,0.50}{##1}}}
\expandafter\def\csname PY@tok@cp\endcsname{\def\PY@tc##1{\textcolor[rgb]{0.74,0.48,0.00}{##1}}}
\expandafter\def\csname PY@tok@gi\endcsname{\def\PY@tc##1{\textcolor[rgb]{0.00,0.63,0.00}{##1}}}
\expandafter\def\csname PY@tok@gh\endcsname{\let\PY@bf=\textbf\def\PY@tc##1{\textcolor[rgb]{0.00,0.00,0.50}{##1}}}
\expandafter\def\csname PY@tok@ni\endcsname{\let\PY@bf=\textbf\def\PY@tc##1{\textcolor[rgb]{0.60,0.60,0.60}{##1}}}
\expandafter\def\csname PY@tok@nl\endcsname{\def\PY@tc##1{\textcolor[rgb]{0.63,0.63,0.00}{##1}}}
\expandafter\def\csname PY@tok@nn\endcsname{\let\PY@bf=\textbf\def\PY@tc##1{\textcolor[rgb]{0.00,0.00,1.00}{##1}}}
\expandafter\def\csname PY@tok@no\endcsname{\def\PY@tc##1{\textcolor[rgb]{0.53,0.00,0.00}{##1}}}
\expandafter\def\csname PY@tok@na\endcsname{\def\PY@tc##1{\textcolor[rgb]{0.49,0.56,0.16}{##1}}}
\expandafter\def\csname PY@tok@nb\endcsname{\def\PY@tc##1{\textcolor[rgb]{0.00,0.50,0.00}{##1}}}
\expandafter\def\csname PY@tok@nc\endcsname{\let\PY@bf=\textbf\def\PY@tc##1{\textcolor[rgb]{0.00,0.00,1.00}{##1}}}
\expandafter\def\csname PY@tok@nd\endcsname{\def\PY@tc##1{\textcolor[rgb]{0.67,0.13,1.00}{##1}}}
\expandafter\def\csname PY@tok@ne\endcsname{\let\PY@bf=\textbf\def\PY@tc##1{\textcolor[rgb]{0.82,0.25,0.23}{##1}}}
\expandafter\def\csname PY@tok@nf\endcsname{\def\PY@tc##1{\textcolor[rgb]{0.00,0.00,1.00}{##1}}}
\expandafter\def\csname PY@tok@si\endcsname{\let\PY@bf=\textbf\def\PY@tc##1{\textcolor[rgb]{0.73,0.40,0.53}{##1}}}
\expandafter\def\csname PY@tok@s2\endcsname{\def\PY@tc##1{\textcolor[rgb]{0.73,0.13,0.13}{##1}}}
\expandafter\def\csname PY@tok@vi\endcsname{\def\PY@tc##1{\textcolor[rgb]{0.10,0.09,0.49}{##1}}}
\expandafter\def\csname PY@tok@nt\endcsname{\let\PY@bf=\textbf\def\PY@tc##1{\textcolor[rgb]{0.00,0.50,0.00}{##1}}}
\expandafter\def\csname PY@tok@nv\endcsname{\def\PY@tc##1{\textcolor[rgb]{0.10,0.09,0.49}{##1}}}
\expandafter\def\csname PY@tok@s1\endcsname{\def\PY@tc##1{\textcolor[rgb]{0.73,0.13,0.13}{##1}}}
\expandafter\def\csname PY@tok@kd\endcsname{\let\PY@bf=\textbf\def\PY@tc##1{\textcolor[rgb]{0.00,0.50,0.00}{##1}}}
\expandafter\def\csname PY@tok@sh\endcsname{\def\PY@tc##1{\textcolor[rgb]{0.73,0.13,0.13}{##1}}}
\expandafter\def\csname PY@tok@sc\endcsname{\def\PY@tc##1{\textcolor[rgb]{0.73,0.13,0.13}{##1}}}
\expandafter\def\csname PY@tok@sx\endcsname{\def\PY@tc##1{\textcolor[rgb]{0.00,0.50,0.00}{##1}}}
\expandafter\def\csname PY@tok@bp\endcsname{\def\PY@tc##1{\textcolor[rgb]{0.00,0.50,0.00}{##1}}}
\expandafter\def\csname PY@tok@c1\endcsname{\let\PY@it=\textit\def\PY@tc##1{\textcolor[rgb]{0.25,0.50,0.50}{##1}}}
\expandafter\def\csname PY@tok@kc\endcsname{\let\PY@bf=\textbf\def\PY@tc##1{\textcolor[rgb]{0.00,0.50,0.00}{##1}}}
\expandafter\def\csname PY@tok@c\endcsname{\let\PY@it=\textit\def\PY@tc##1{\textcolor[rgb]{0.25,0.50,0.50}{##1}}}
\expandafter\def\csname PY@tok@mf\endcsname{\def\PY@tc##1{\textcolor[rgb]{0.40,0.40,0.40}{##1}}}
\expandafter\def\csname PY@tok@err\endcsname{\def\PY@bc##1{\setlength{\fboxsep}{0pt}\fcolorbox[rgb]{1.00,0.00,0.00}{1,1,1}{\strut ##1}}}
\expandafter\def\csname PY@tok@mb\endcsname{\def\PY@tc##1{\textcolor[rgb]{0.40,0.40,0.40}{##1}}}
\expandafter\def\csname PY@tok@ss\endcsname{\def\PY@tc##1{\textcolor[rgb]{0.10,0.09,0.49}{##1}}}
\expandafter\def\csname PY@tok@sr\endcsname{\def\PY@tc##1{\textcolor[rgb]{0.73,0.40,0.53}{##1}}}
\expandafter\def\csname PY@tok@mo\endcsname{\def\PY@tc##1{\textcolor[rgb]{0.40,0.40,0.40}{##1}}}
\expandafter\def\csname PY@tok@kn\endcsname{\let\PY@bf=\textbf\def\PY@tc##1{\textcolor[rgb]{0.00,0.50,0.00}{##1}}}
\expandafter\def\csname PY@tok@mi\endcsname{\def\PY@tc##1{\textcolor[rgb]{0.40,0.40,0.40}{##1}}}
\expandafter\def\csname PY@tok@gp\endcsname{\let\PY@bf=\textbf\def\PY@tc##1{\textcolor[rgb]{0.00,0.00,0.50}{##1}}}
\expandafter\def\csname PY@tok@o\endcsname{\def\PY@tc##1{\textcolor[rgb]{0.40,0.40,0.40}{##1}}}
\expandafter\def\csname PY@tok@kr\endcsname{\let\PY@bf=\textbf\def\PY@tc##1{\textcolor[rgb]{0.00,0.50,0.00}{##1}}}
\expandafter\def\csname PY@tok@s\endcsname{\def\PY@tc##1{\textcolor[rgb]{0.73,0.13,0.13}{##1}}}
\expandafter\def\csname PY@tok@kp\endcsname{\def\PY@tc##1{\textcolor[rgb]{0.00,0.50,0.00}{##1}}}
\expandafter\def\csname PY@tok@w\endcsname{\def\PY@tc##1{\textcolor[rgb]{0.73,0.73,0.73}{##1}}}
\expandafter\def\csname PY@tok@kt\endcsname{\def\PY@tc##1{\textcolor[rgb]{0.69,0.00,0.25}{##1}}}
\expandafter\def\csname PY@tok@ow\endcsname{\let\PY@bf=\textbf\def\PY@tc##1{\textcolor[rgb]{0.67,0.13,1.00}{##1}}}
\expandafter\def\csname PY@tok@sb\endcsname{\def\PY@tc##1{\textcolor[rgb]{0.73,0.13,0.13}{##1}}}
\expandafter\def\csname PY@tok@k\endcsname{\let\PY@bf=\textbf\def\PY@tc##1{\textcolor[rgb]{0.00,0.50,0.00}{##1}}}
\expandafter\def\csname PY@tok@se\endcsname{\let\PY@bf=\textbf\def\PY@tc##1{\textcolor[rgb]{0.73,0.40,0.13}{##1}}}
\expandafter\def\csname PY@tok@sd\endcsname{\let\PY@it=\textit\def\PY@tc##1{\textcolor[rgb]{0.73,0.13,0.13}{##1}}}

\def\PYZbs{\char`\\}
\def\PYZus{\char`\_}
\def\PYZob{\char`\{}
\def\PYZcb{\char`\}}

\def\PYZsh{\char`\#}

\def\PYZdl{\char`\$}
\def\PYZhy{\char`\-}
\def\PYZsq{\char`\'}
\def\PYZdq{\char`\"}

\makeatother

    \definecolor{incolor}{rgb}{0.0, 0.0, 0.5}
    \definecolor{outcolor}{rgb}{0.545, 0.0, 0.0}

\newcommand{\figurefolder}{figures}
\newcommand{\apxfolder}{apx/QInfer Tomography Tutorial_files}

\newcommand{\cj}{Choi-Jamio{\l}kowski}
\newcommand{\sample}{\mathrm{sample}}
\newcommand{\etal}{\emph{et al.}}

\newcommand{\T}{\mathrm{T}}
\newcommand\Tr{\mathrm{Tr}}
\newcommand\supp{\operatorname{supp}}
\newcommand\Cov{\operatorname{Cov}}

\newcommand{\CC}{\mathbb{C}}
\newcommand{\expect}{\mathbb{E}}
\newcommand{\Var}{\mathbb{V}}
\newcommand{\id}{\mathbbm{1}}

\newcommand{\diag}{\operatorname{diag}}

\newcommand{\ii}{\mathrm{i}}
\newcommand{\dd}{\mathrm{d}}

\newcommand{\Nor}{\mathrm{N}}
\newcommand{\Like}{\mathcal{L}}
\newcommand{\Id}{\openone}

\newcommand{\expt}[1]{\langle {#1} \rangle}
\newcommand{\sket}[1]{\left | {#1} \right\rrangle}
\newcommand{\sbra}[1]{\left\llangle {#1} \right |}
\newcommand{\sbraket}[1]{\left\llangle {#1} \right\rrangle}
\newcommand{\sip}[2]{\left\llangle {#1}| {#2} \right\rrangle}

\newcommand{\defeq}{\mathrel{:=}}
\newlength{\dhatheight}

\makeatletter
\def\Decl@Mn@Delim#1#2#3#4{%
  \if\relax\noexpand#1%
    \let#1\undefined
  \fi
  \DeclareMathDelimiter{#1}{#2}{#3}{#4}{#3}{#4}}
\def\Decl@Mn@Open#1#2#3{\Decl@Mn@Delim{#1}{\mathopen}{#2}{#3}}
\def\Decl@Mn@Close#1#2#3{\Decl@Mn@Delim{#1}{\mathclose}{#2}{#3}}
\DeclareFontFamily{OMX}{MnSymbolE}{}
\DeclareFontShape{OMX}{MnSymbolE}{m}{n}{
    <-6>  MnSymbolE5
   <6-7>  MnSymbolE6
   <7-8>  MnSymbolE7
   <8-9>  MnSymbolE8
   <9-10> MnSymbolE9
  <10-12> MnSymbolE10
  <12->   MnSymbolE12}{}
\DeclareFontShape{OMX}{MnSymbolE}{b}{n}{
    <-6>  MnSymbolE-Bold5
   <6-7>  MnSymbolE-Bold6
   <7-8>  MnSymbolE-Bold7
   <8-9>  MnSymbolE-Bold8
   <9-10> MnSymbolE-Bold9
  <10-12> MnSymbolE-Bold10
  <12->   MnSymbolE-Bold12}{}
\DeclareSymbolFont{mnsymbols}  {OMX}{MnSymbolE}{m}{n}
\Decl@Mn@Open {\lsem}               {mnsymbols}{'102}
\Decl@Mn@Close{\rsem}               {mnsymbols}{'107}
\Decl@Mn@Open {\llangle}            {mnsymbols}{'164}
\Decl@Mn@Close{\rrangle}            {mnsymbols}{'171}
\makeatother

\newcommand{\eq}[1]{\hyperref[eq:#1]{(\ref*{eq:#1})}}
\renewcommand{\sec}[1]{\hyperref[sec:#1]{Section~\ref*{sec:#1}}}
\newcommand{\apx}[1]{\hyperref[apx:#1]{Appendix~\ref*{apx:#1}}}
\newcommand{\tab}[1]{\hyperref[tab:#1]{Table~\ref*{tab:#1}}}
\newcommand{\alg}[1]{\hyperref[alg:#1]{Algorithm~\ref*{alg:#1}}}
\newcommand{\vid}[1]{\hyperref[vid:#1]{Video~\ref*{vid:#1}}}
\newcommand{\fig}[1]{\hyperref[fig:#1]{Figure~\ref*{fig:#1}}}
\newcommand{\thm}[1]{\hyperref[thm:#1]{Theorem~\ref*{thm:#1}}}
\newcommand{\lem}[1]{\hyperref[lem:#1]{Lemma~\ref*{lem:#1}}}
\newcommand{\cor}[1]{\hyperref[cor:#1]{Corollary~\ref*{cor:#1}}}
\newcommand{\defn}[1]{\hyperref[def:#1]{Definition~\ref*{def:#1}}}

\usepackage[breaklinks=true]{hyperref}
\hypersetup{
  colorlinks   = true, 
  urlcolor     = blue, 
  linkcolor    = blue, 
  citecolor   = red 
}

\usepackage{algorithm} 

\newcommand{\newaffil}[2]{
    \expandafter\newcommand\csname affil#1\endcsname{
        \affiliation{
            #2
        }
    }
}
\newaffil{TODO}{\TODO}
\newaffil{EQUS}{
    Centre for Engineered Quantum Systems,
    University of Sydney,
    Sydney, NSW, Australia
}
\newaffil{USydPhys}{
    School of Physics,
    University of Sydney,
    Sydney, NSW, Australia
}
\newaffil{IQC}{
    Institute for Quantum Computing,
    University of Waterloo,
    Waterloo, ON, Canada
}
\newaffil{UWAMath}{
    Department of Applied Mathematics,
    University of Waterloo,
    Waterloo, ON, Canada
}
\newaffil{UWChem}{
    Department of Chemistry,
    University of Waterloo,
    Waterloo, ON, Canada
}
\newaffil{PI}{
    Perimeter Institute for Theoretical Physics, 31 Caroline St. N, Waterloo, Ontario, Canada N2L 2Y5
    }
\newaffil{CIFAR}{
    Canadian Institute for Advanced Research,
    Toronto, ON, Canada
}

\begin{document}


\title{Practical Bayesian Tomography}

\author{Christopher Granade}
\email[]{cgranade@cgranade.com}
\homepage[]{http://cgranade.com/}
\thanks{\\
    Literate source code for the figures, animations and tutorials
    appearing in this work is available at \url{http://goo.gl/fRqnIn}.
}
\affilEQUS
\affilUSydPhys

\author{Joshua Combes}
\affilIQC
\affilUWAMath
\affilPI

\author{D. G. Cory}
\affilIQC
\affilPI
\affilUWChem
\affilCIFAR

\date{\today}



\begin{abstract}
    In recent years, Bayesian methods have been proposed as a solution to a
    wide range of issues in quantum state and process tomography. 
    State-of-the-art 
    Bayesian tomography solutions suffer from three problems: numerical intractability, 
    a lack of informative prior distributions, and an inability to track time-dependent 
    processes. Here, we address all three problems. First, we use modern statistical methods, 
    as pioneered by Husz\'ar and Houlsby \cite{HH12} and by Ferrie \cite{ferrie_quantum_2014}, 
    to make Bayesian tomography numerically tractable. Our approach allows for practical 
    computation of Bayesian point and region estimators for quantum states and channels. 
    Second, we propose the first priors on quantum states and channels that allow
    for including useful experimental insight. 
    Finally, we develop a method that allows tracking of time-dependent states
    and estimates the drift and diffusion processes affecting a state.
    We provide source code and animated visual examples for our methods.
\end{abstract}


\maketitle

\section{Introduction}
\label{sec:intro}

Quantum state and process tomography are important methods for diagnosing and
characterizing imperfections in small quantum systems. By fixing problems in
models and implementations, and by
having a well-characterized system, we may hope to compose multiple systems to
build reliable larger quantum systems. These larger systems require a scalable
approach to characterization, such as using the matrix-product state ansatz
\cite{cramer_efficient_2010} or information locality
\cite{wiebe_quantum_2015,holzapfel_scalable_2015}. Quantum tomography has seen
many improvements since its inception
\cite{NewtonYoung68,*BandPark70,*BandPark79}. In particular, tomography has
enjoyed advances in providing maximum-likelihood estimators \cite{Hradil97},
region estimators \cite{christandl_reliable_2011,RBK12,shang_optimal_2013},
model selection \cite{guta_rank-based_2012,enk_when_2013}, hedging 
\cite{blume-kohout_hedged_2010}, and compressed sensing
\cite{gross_quantum_2010,flammia_quantum_2012}.

These techniques, though powerful, do not take advantage of prior information
available to experimentalists. Such prior information can include knowledge
gained in building the experiment, or in performing similar experiments, as
well as knowledge gained from the calibration leading up to an experiment of
interest. A class of techniques that allow one to include prior information is
called Bayesian estimation.

Bayesian techniques in the context of quantum
tomography were first suggested by Jones \cite{Jones91a,Jones91b,Jones94},
Slater \cite{Slater95}, Derka \etal ~\cite{DerBuzAdmKni96}, Bu\v{z}ek
\etal~\cite{BuzDerAda98}, and Schack \etal~\cite{SchBruCav01}.
In addition to the inclusion of prior information,
Bayesian estimation also naturally includes several other experimental
advantages, such as optimality \cite{HadBlum06, RBK10,FerKue15}, adaptive experimental design \cite{HH12,kravtsov_experimental_2013}, robust region
estimates \cite{ferrie_high_2014} and model selection criteria
\cite{ferrie_quantum_2014,wiebe_quantum_2014}.
These advantages arise from the fact that Bayesian methods provide a complete
characterization of the current state of an experimentalist's knowledge after
each datum. 

Given the many proposals for and advantages of Bayesian methods, the lack of adoption of Bayesian methods in experimental tomography, with the exception of some recent work
~\cite{kravtsov_experimental_2013,struchalin_experimental_2015}, seems to be 
primarily a practical problem. Bayesian methods are rarely analytically tractable. Further 
numerical implementations of Bayesian methods (for inference of many variables) are a 
small scale software engineering project and computationally expensive. Bayesian reasoning, in the classical statistics literature,  is typically 
realized with efficient well-known classical algorithms which go
by the names of particle filtering \cite{DouJoh09a} and sequential Monte Carlo
\cite{doucet_sequential_2000}. Even though these algorithms can be numerically efficient,
for multiple variables they are non-trivial to implement and optimize. In the context of quantum state tomography 
sequential Monte Carlo has been applied to adaptive tomography \cite{HH12} and model selection \cite{ferrie_quantum_2014}, generalizing and dramatically simplifying earlier efforts based on Kalman filtering \cite{audenaert_quantum_2009}.
However, much of the code developed for this application is either specialized to particular cases,
or has not been released to or adopted by the community. Releasing reusable code is critical not only for practicality in experiments, but also
for producing reproducible research results \cite{ince_case_2012}.

Another difference between the application of Bayesian methods in classical statistics and its application
to quantum state and processes tomography is the lack of choice in priors. Experimentalists spend many hours designing, testing and calibrating their platforms. It would be nice to include the prior information from this lengthy process into a quantum estimation procedure. Classically one can choose from many different priors, which could be ``informative'' or ``uninformative,''  depending on the domain of the probability distribution. In the quantum setting, many canonical priors are unitarily invariant and have some uninformative distribution over purity \cite{BorNad12}. The lack of alternatives can be explained both by the lack of theoretical knowledge on how to construct such priors and more importantly there is no general purpose software available that could make use of such priors once constructed.

Finally, there is a critical issue facing the tomography community, independent of the respective merits
of frequentist or Bayesian approaches. It is the fact that the output of realistic quantum sources
vary in time, both deterministically (``drift'') and stochastically (``diffusion''). Incorporating time-dependence in quantum characterization protocols has been a topic of
significant recent interest \cite{schwarz_detecting_2011,enk_when_2013,Lan13,shulman_suppressing_2014, fogarty_nonexponential_2015, granade_characterization_2015}.
For the most part, model selection has been used in tomographic experiments to detect
drift or diffusion.
While useful, this does not yield a protocol for incorporating that drift (or diffusion) into the estimation protocol
once it has been detected.
Moreover, many of the current solutions are not general purpose, nor are they
practical in a range of important applications.
For instance, see the proposal
of Blume-Kohout \etal~\cite{[{The third paragraph on page 22 of }]blume2015turbocharging}, which
requires quantum memory on the order of the number of samples collected so that the DeFinetti conditions are met. This is clearly impractical, and further precludes adaptivity.

In this work, we address all three issues. The unifying theme is the use of the sequential Monte Carlo (SMC)
algorithm to perform tomography. 

To address the issue of adoption we implement SMC-based tomography using
an open-source library for Python \cite{granade_qinfer:_2012} that integrates
with the widely-used QuTiP library for quantum information
\cite{johansson_qutip_2013}, and can be used with modern instrument control
software \cite{casagrande_instrumentkit:_2013}.  The techniques that we
introduce in this paper can readily be applied in experimental practice---we 
provide a tutorial on software implementations in \apx{qinfer-tutorial}.
    
Second, we give a constructive method for defining priors that
represent initial estimates informed by experimental insight.
We use sequential Monte Carlo to avoid the need to write
an analytic form for our prior. This is especially useful in the context
of quantum state and process tomography, where analytic expressions are
available for only a few distributions.
Instead, SMC only requires that we provide a means to sample
from priors.
To do so, we first draw sample states or channels from a
reference or ``flat'' prior, then transform each sample by a quantum
operation drawn from an ensemble. The only input required to define this
ensemble is a prior estimate of the state. Our method then guarantees that
this becomes the mean of our new prior.

Third, we incorporate tracking of drifting and diffusing evolution into state
tomography by using techniques from computer vision which perturb a sample. Each such perturbation is
drawn from a Gaussian whose mean and variance represent deterministic and
stochastic evolution respectively.

This article is structured as follows. We begin by reviewing Bayesian
tomography in \sec{bayesian-tomography}. Then we describe prior work on
priors for quantum states and channels in \sec{uninformative-priors}. 
In \sec{informative-priors} we transform these priors into
new priors for states and channels that include experimental insight.
Using Bayesian tomographic methods, we then
describe how to track quantum states and channels as a function of time in
\sec{state_space_tracking}. In \sec{examples} we illustrate these ideas and
more, then conclude in \sec{conclusion}.

\section{Bayesian Tomography}
\label{sec:bayesian-tomography}

To infer the state of a quantum system of dimension $D$, we perform a set of
measurements on $N$ identically and independently (iid) prepared quantum
systems. As is usually the case, we will restrict to measuring each system
separately. In general, one could consider performing a different generalized
measurement on each system, and the kind of generalized measurement could
adaptively depend on all prior measurements. This can be included in the
expressions below at the cost of additional notational baggage.

Consider a single positive operator valued measure (POVM) $\{E_k\}$ whose $K$ elements represent the outcomes of measurements. If we perform this generalized measurement on all $N$ systems it results in a string of measurement results $\mathcal M =\{ M_{1}, M_{2}, ... , M_{N} \}$, where $M_{i}$ is the POVM element obtained in the $i$th trial. Let $n_k$ be the number times that $E_{k}$ is observed in
$\mathcal M$, i.e. frequency of $E_{k}$. The statistical information measurement outcomes have about the preparation
is described by a \emph{likelihood function}; that is, a probability distribution over measurement
records, conditioned on a hypothesis $\rho$ about the state. In particular, by using Born's rule
$\Pr(E_k | \rho) = \Tr(E_k \rho)$ to write out the likelihood, we obtain that
\begin{subequations}
    \begin{align}
        \Like(\rho)  &= \Pr(\mathcal M| \rho)
          = \prod_{i=1} ^N \Tr \left [M_{i} \rho\right ] \label{eq:likelihood1}\\
        & =  \Tr \left [E_{1} \rho\right ]^{n_1}\Tr \left [E_2 \rho\right ]^{n_2}\ldots \Tr \left [E_K \rho\right ]^{n_K} \label{eq:likelihood2}
    \end{align}
\end{subequations}
Moreover, by using the \cj\ isomorphism,
we can associate a state $J(\Lambda)/D$ with each quantum channel $\Lambda$.
As we detail in \apx{choi}, preparation and measurement in a process tomography
experiment can be written as a measurement of the state $J(\Lambda)/D$,
such that \eq{likelihood1} also includes this case.

In general, a likelihood function completely models an experiment by specifying the
probability of observing any measurement, conditioned on the hypothesis we
would like to learn. Thus, the likelihood function serves as the
basis for subsequent estimation. 

For instance, in maximum likelihood estimation (MLE),
an estimate $\hat{\rho}$ of the state $\rho$ is formed by
\begin{align}
        \hat{\rho}_{\text{MLE}} & \defeq \underset{\rho}{\operatorname{arg\,max}}\,\Like(\rho).
\end{align}
Here, however, we use the likelihood function to instead
perform Bayesian inference as described by Jones \cite{Jones91a,*Jones91b,*Jones94}, Slater \cite{Slater95}, Derka \etal ~\cite{DerBuzAdmKni96}, Bu\v{z}ek \etal \cite{BuzDerAda98}, and Schack \etal~\cite{SchBruCav01}. Below we follow Blume-Kohout's \cite{RBK10} presentation closely.
We begin by using Bayes' rule,
\begin{align}
    \Pr(\rho | \mathcal{M})\,\dd\rho = \frac{\Pr(\mathcal M| \rho) \Pr(\rho)\,\dd\rho}{ \Pr(\mathcal{M}) }, 
\end{align}
where $\pi(\rho) \defeq \Pr(\rho) \dd\rho$ is called the \emph{prior} distribution,
and 
\begin{align}
    \mathcal{N} = \Pr(\mathcal{M}) = \int\dd\rho\Pr(\mathcal M| \rho) \Pr(\rho)
\end{align}
is a normalization constant. We shall write $\rho \sim \pi$ to indicate
that the random variable $\rho$ is drawn from the prior $\pi$.

In the next two Sections, we will return to the question of how to choose the
prior distribution $\Pr(\rho)\dd \rho$. Henceforth, we drop the measure $\dd \rho$
unless we are integrating a distribution, as we will later use a numerical
algorithm which approximates this continuous distribution by a discrete
distribution.

The Bayesian mean estimate (BME) is then given by the expectation
\begin{align}
    \label{eq:rho_est}
    \hat{\rho}_{\text{BME}}(\mathcal M)
        & \defeq \expect_{\rho}[\rho | \mathcal{M}]
          = \int \dd\rho\, \rho \Pr(\rho | \mathcal{M}) ,
\end{align}
where $\expect$ indicates an expectation value, and where conditional bars
and the subscript denote the distribution the expectation is taken over e.g. 
$\expect_{x}[f(x)|y] = \sum_x f(x) \Pr(x|y) $ denotes the conditional expectation 
of $f(x)$ given $y$.

The Bayesian mean estimator is an optimal estimator
for any strictly proper scoring rule on states
\cite{HadBlum06}. These scoring rules arise from Bregman divergences \cite{banerjee_optimality_2005}
such as the Kullback-Leibler divergence or the quadratic loss
$L_Q(\rho, \hat{\rho}) \defeq \Tr\left[(\rho - \hat{\rho}) Q [\rho - \hat{\rho}]\right]$, where $Q$
is a positive semidefinite superoperator
\footnote{This definition of the quadratic loss follows from the typical
definition by interpreting the scale matrix as a superoperator.}.
As we will discuss in more detail below, the error incurred by the BME is
well-characterized by spread of samples from the posterior distribution.
Importantly, if one uses infidelity as a loss function, the BME remains
approximately optimal, even though the infidelity is not a Bregman divergence
rule~\cite{FerKue15}.

To make the problem of estimating states and channels more concrete, it is helpful
to specify a real-valued parameterization of the tomographic model. We start
by considering
the space of linear operators acting on a $D$-dimensional Hilbert
space. We represent an operator with an ``operator ket'' $\sket{A}$
corresponding to $A$, while the dual vector is the corresponding ``bra''
$\sbra{A}$ and represents $A^\dagger$. This vector space has the Hilbert-Schmidt
inner product  $\sip{A}{B} = \Tr[A^\dag B]$. In the $D$-dimensional Hilbert
space, a state matrix can be represented as
\begin{align}
    \rho =  \frac{1}{D} \id + \sum_{\alpha=1}^{D^2-1}x_\alpha B_\alpha
\end{align}
where $x_\alpha = \sbraket{B_\alpha | \rho}$ for a basis of Hermitian operators
$\{ B_\alpha \}$ that is orthonormal under the Hilbert-Schmidt inner product
$\sip{B_\alpha}{B_\beta} = \Tr[B_\alpha^\dagger B_\beta]=\delta_{\alpha,\beta}$. For simplicity, we choose
$B_0 = \id / \sqrt {D}$ to be the only traceful element.
The corresponding operator ket representation of $\rho$ is
\begin{align}
    \sket{\rho} = \left(\begin{array}{c}
        x_0 \\
        x_1 \\ x_2 \\ \vdots \\ x_{D^2-1}
    \end{array}\right) = \left(\begin{array}{c}
        \sip{B_0}{\rho} \\
        \sip{B_1}{\rho} \\ \sip{B_2}{\rho} \\ \vdots \\ \sip{B_{D^2-1}}{\rho}
    \end{array}\right) = \left(\begin{array}{c}
        1 / \sqrt{D} \\
        \Tr[B_1\rho] \\ \Tr[B_2\rho] \\ \vdots \\ \Tr[B_{D^2-1}\rho]
    \end{array}\right).
\end{align}
As a consequence of $\rho$ being considered as a random variable, 
each parameter $x_i$ is also a random variable. That is, we have represented
$\rho$ in terms of a vector-valued random variable $\sket{\rho}$;
because we have chosen a Hermitian basis, $\sket{\rho}$ is also a real vector.
Moreover, each parameter $x_i$ is then by definition equal to the
mean value $\expt{B_i}$ of the \emph{observable} $B_i$, taken over possible
measurement outcomes.
In general, tensor products of Pauli matrices and generalized Gell-Mann
matrices can be used as such a Hermitian basis for multiple finite-dimensional
quantum systems.

Having chosen a parameterization in terms of observables, we can
now reason about the \emph{error} in parameters of $\hat\rho$.
Suppose that for a given posterior, $\rho$ is normally-distributed about the estimated state $\hat{\rho}$,
then the distribution over $\rho$ is fully described by its mean, i.e. \eq{rho_est}, and covariance $\sigma_{i,j}$
between $\hat{x}_i$ and $\hat{x}_j$ (the $i$ and $j$ components of $\hat\rho$),
\begin{align}
    \sigma_{i,j} = \Cov_{i,j}(\sket{\rho}) = \expect_\rho[(x_i-\expect_\rho[ x_i ])(x_j-\expect_\rho[ x_j ])]. 
\end{align}
The covariance matrix for the posterior is then
\begin{gather}
\begin{aligned}
    \Sigma \rho &=
    \expect_\rho [\sket{\rho - \expect[\rho]}\sbra{\rho - \expect[\rho]}]
    = \Cov(\sket{\rho}) \\
&=    \left(\begin{array}{ccccc}
        0 & 0 & 0 & \ldots & 0 \\
        0 & \sigma_{1,1} & \sigma_{1,2} &\ldots & \sigma_{1, D^2-1} \\
        0 & \sigma_{1,2} & \sigma_{2,2} &   & \sigma_{2, D^2-1} \\
        0 & \vdots &   & \ddots & \vdots \\
        0 & \sigma_{1, D^2-1} & \sigma_{2, D^2-1} &\ldots & \sigma_{D^2-1, D^2-1}
    \end{array}\right).
\end{aligned}
\end{gather}
Because we have written the covariance matrix in the basis of $\sket{\rho}$,
we can apply $\Sigma\rho$ as a superoperator on linear operators. The action
of $\Sigma\rho$ on an observable $X$ then gives the variance of the value taken
on by measurements of $X$ in terms of the law of total variance as
\begin{gather}
    \begin{aligned}
        \Var[X] & = \Var_{\rho}[\expt{X}_\rho] + \expt{X^2}_{\expect[\rho]} - \expt{X}^2_{\expect[\rho]} \\
                & = \sbra{X}\Sigma\rho\sket{X} + \sbra{X} X\sket{\hat\rho} - \expt{X}^2_{\expect[\rho]},
    \end{aligned}
\end{gather}
where $\Var_{\rho}$ is the variance over the state $\rho$ and where $\langle\cdot\rangle_\rho$
is the expectation value over measurement outcomes, conditioned on the state $\rho$.
Note that although we have used a coordinate form to arrive at the above expression,
it is basis independent.

As discussed in detail by Blume-Kohout~\cite{RBK10}, the covariance matrix
$\Delta \rho$ describes a credible region (ellipsoid) up to a scaling parameter
$Z$ corresponding to the level of the region \cite{granade_robust_2012}. Specifically, the
eigenvectors and eigenvalues of $\Delta \rho$ are the principle axes and
lengths of the axes respectively. Minimizing an appropriate norm of
$\Sigma\rho$ thus provides a natural objective function for adaptively designing 
tomographic experiments, as we will discuss further in \sec{examples}.

Returning to the problem of finding posteriors, we note that in practice, the
integral in \eq{rho_est} is rarely analytically tractable. Thus, Blume-Kohout
suggested an approximation such as the Metropolis-Hastings algorithm could be
used instead \cite{RBK10}. Rejection sampling methods such as
Metropolis-Hastings tend to be prohibitively expensive, however, and suffer from
vanishingly small acceptance probabilities as data is collected. Though there
have been recent advances in rejection sampling for Bayesian inference
\cite{wiebe_rejection_2015}, the assumption of a normal posterior is difficult to use in the
context of quantum tomography. Consequently, we instead follow the approach of Husz\'ar and
Houlsby \cite{HH12}, and later Ferrie \cite{ferrie_high_2014}, and use
the sequential Monte Carlo (SMC) algorithm \cite{doucet_sequential_2000},
which does not rely on the assumption of a normal posterior.
A brief review of SMC can be found in \apx{SMC} and the
references therein. 

SMC offers the advantage that we need not explicitly write down a prior, but
instead treat Bayes' rule as a transition kernel that transforms prior
hypotheses called \emph{particles} \cite{doucet_tutorial_2011} into
samples from the posterior. These particles are then used to approximate the
integral \eq{rho_est}. For tomography, each particle represents a particular
hypothesis about the state $\rho$, so that the prior $ \Pr(\rho) \dd\rho$ can be written under the SMC approximation as
\cite{HH12}
\begin{align}
    \label{eq:SMC_prior} 
    \Pr(\rho) \approx \sum_{p\in \text{particles}} w_p
        \, \delta(\rho -\rho_p),
\end{align}
where $w_p$ reflects the relative plausibility of the corresponding state
conditioned on all available evidence. Initially, $w_p$ is taken to be uniform,
as the density of samples carries information about prior plausibility. After
updating the particles based on experimental observations, we can readily
calculate the Bayesian mean estimator and posterior covariance matrix by
summing over the particle approximation.

Before moving on, we wish to point out that SMC also allows for more sophisticated credible region
estimators--- we focus here on the covariance region estimator for simplicity.
In particular, any set of particles $P_\alpha$ such that
$\sum_{p\in P_\alpha} w_p \ge 1 - \alpha$ forms an $\alpha$-credible region.
Taking a convex hull over such a region then provides a region which naturally
includes the convexity of state space, and a minimum-volume enclosing
ellipsoid over a credible region yields a compact description \cite{ferrie_high_2014}.
Both of these credible region estimators are included in the open-source package
that we rely on for numerical implementations, QInfer \cite{granade_qinfer:_2012}.
Thus, we inherit a variety of practical data-driven credible region estimators.
For the 
remainder of this work, we choose to use to use naive ``$3\sigma$'' covariance
ellipsoids for the purpose of illustration,
by which we mean that we take the ellipsoid defined by the covariance matrix
and scale it by a factor of $Z=3$.

\section{Default Priors: the Sampling of States and Channels}
\label{sec:uninformative-priors}

In sequential Monte Carlo, we need to be able to draw samples from a prior, see \eq{SMC_prior}. In
this Section we briefly review how to draw samples from several
well-established priors \cite{Jones91a,*Jones91b,*Jones94}.
Loosely speaking, these priors are useful as they define a notion of uniformity
over states and channels, and do not posit any prior estimate other than the maximally
mixed state.
Following the advice of Wasserman \cite{Wasserman_priors}, we will term these
well-established priors as \emph{default priors}, as each of them is a reasonable
choice to adopt as a prior in lieu of more detailed information.

Later, in \sec{informative-priors}, we will take priors
as algorithmic \emph{inputs} that define what states are feasible for a given
tomographic experiment. We will refer to priors that can be used in this way
as \emph{fiducial}.
From the perspective of the assumptions our algorithm makes,
we require inputs to have the property that
\begin{align}\label{eq:uninformed}
    \expect_{\rho\sim\pi}[\rho] =& \int \rho\,\pi(\rho)=\int \rho \Pr(\rho) \dd\rho = \frac\Id D; \\
    & ({\rm fiducial\ prior}) \nonumber
\end{align}
that is, that the mean of the prior is the maximally mixed state. 
All of the default priors described in this section are fiducial in the sense given in \eq{uninformed}.
Similarly, we say that a prior is \emph{insightful} if its mean is anything
other than a maximally-mixed state. 
In \sec{informative-priors}, we consider priors that are insightful by our definition, that is 
\begin{align}\label{eq:informed}
    \expect_{\rho\sim\pi}[\rho] =& \int \rho\,\pi(\rho)=\int \rho \Pr(\rho) \dd\rho = \rho_\mu, \\
    & ({\rm insightful\ prior}) \nonumber
\end{align}
where $\rho_\mu \ne \id / D$.

In constructing default priors, we will make repeated use of random
complex-valued matrices with entries sampled from normal distributions. Such
matrices form the Ginibre matrix ensemble \cite{OsiSomZyc10}. We provide pseudocode
for all default sampling algorithms in \apx{sampling-algorithms}. For brevity, we refer to algorithms
by the initials of their authors; for instance, we refer to \alg{ginibre-matrix}
as the ZS algorithm after \.{Z}yczkowski and Sommers \cite{zyczkowski_induced_2001}.

\subsection{Priors on States}

For pure quantum states, the canonical default prior is the Haar
measure. One can easily sample states from this measure by sampling a vector
in $\CC^D$ with Gaussian-distributed entries, then renormalizing.
Alternatively, one can sample unitary matrices uniformly
according to the Haar measure as detailed in the Mezzardi algorithm
\cite{mezzadri_how_2007}, see \alg{haar-unitary}. A random pure state is then a
Haar-random unitary applied to a fiducial state. Note that the Haar measure is
\emph{fiducial}; that is, it makes a prediction of the maximally-mixed state,
in the sense of \eq{uninformed}.

Generalizing to mixed states, we consider two well-known ensembles of random states.
First, we consider states drawn from the Ginibre ensemble, a generalization of
the Hilbert-Schmidt ensemble that allows for restrictions on rank.
Second, we consider the ensemble of states drawn from the Bures measure.

Samples from both ensembles are neatly captured by a single equation
\begin{align}
    \label{eq:ginibre-bures}
    \rho_{\rm sample} = \frac{(\Id + U)AA^\dag(\Id+U^\dag)}{\Tr[(\Id + U)AA^\dag(\Id+U^\dag)]},
\end{align}
where $U$ is either the identity or a Haar-random unitary, and $A$ is a $D\times K$ Ginibre matrix. If
$U$ is taken to be the identity, then the state is drawn from the Ginibre
ensemble with rank $K$ and is unitarily invariant. This means the prior will only have support on states with rank less than or equal to $K$.  If $A$ is taken to be a
$D\times D$ Ginibre matrix and $U$ is Haar-random, then the state is drawn from
the Bures measure. Thus, $\rho_{\rm sample} \sim \operatorname{Ginibre}(D, K)$ or
$\rho_{\rm sample} \sim \operatorname{Bures}(D)$, respectively. These samples can then serve as an fiducial prior for SMC, in the sense described by equations \eq{SMC_prior} and \eq{uninformed}, or can be transformed into samples from a prior that is insightful, as described in \sec{informative-priors}.

These procedures are given by \alg{ginibre-state} and \alg
{bures-state} respectively \footnote{More generally, Osipov \etal~\cite{OsiSomZyc10} show that
by linearly interpolating between $\id$ and $U$ in \eq{ginibre-bures},
one obtains a continuous family of distributions with the Hilbert-Schmidt and
Bures ensembles as its extrema.}.

\subsection{Priors on Channels}

In developing applications to quantum process tomography (QPT), we use the fact that learning
the Choi states of unknown channels is a special case of state tomography, as derived in \apx{choi}.
Thus, it is also useful to
consider prior distributions over the Choi states of completely positive trace preserving (CPTP)
quantum maps. In particular, for process tomography, we will use the measure derived by 
BCSZ \cite{bruzda_random_2009} to draw samples from a prior over quantum channels that is fiducial.
The resulting algorithm is unitarily invariant and supported
over all channels of a given Kraus rank; that is the minimal number of Kraus 
operators required to specify the channel.

As detailed by Bruzda \etal~\cite{bruzda_random_2009}, to generate a channel $\Lambda : \CC^D \to \CC^D$,
the BCSZ algorithm (see \alg{random-cptp-map}) begins by selecting a Ginibre-random density operator $\rho$ of dimension
$D^2$ and a fixed rank $K$. For notational simplicity, we let $\rho$ be an operator
acting on the bipartite Hilbert space $\CC^D \otimes \CC^D$.
The trace-preserving condition is then imposed by letting $Y = \Tr_2 \rho$ be the
partial trace over the second copy of $\CC^D$, then transforming
$\rho$ into the Choi state of the sampled channel $\Lambda_\sample$ by
\begin{equation}
    J(\Lambda_\sample) / D = (Y^{-1/2} \otimes \id) \rho (Y^{-1/2} \otimes \id).
\end{equation}
It is easy to verify that channels $\Lambda_\sample$ sampled in this way are indeed
trace-preserving and completely positive. Moreover, the transformation
above preserves the property that $\expect[J(\Lambda_\sample) / D] = \id / D$,
such that the mean of the BCSZ distribution is the completely depolarizing
channel. This condition on channel priors is precisely that given as \eq{uninformed}. 
We will show below that the BCSZ distribution suffices to construct a
prior over channels that is insightful.

\section{ Insightful Priors for States and Channels}
\label{sec:informative-priors}

Our basic technique is to transform the samples drawn from fiducial priors,
in particular the default priors described in \sec{uninformative-priors},
to insightful priors by applying a channel $ \Phi$ to the fiducial prior
\begin{align}
    \underbrace{    \rho_{\rm sample}}_{\rm insightful}=     \Phi \big ( \!\! \underbrace{\rho_{\rm sample}}_{\quad\rm fiducial\quad} \!\!\big ).
\end{align}
The algebraic gymnastics below simply determine how to construct insightful priors with a given mean $\rho_\mu$.

\begin{figure}
    \begin{centering}
       \includegraphics[width=\columnwidth]{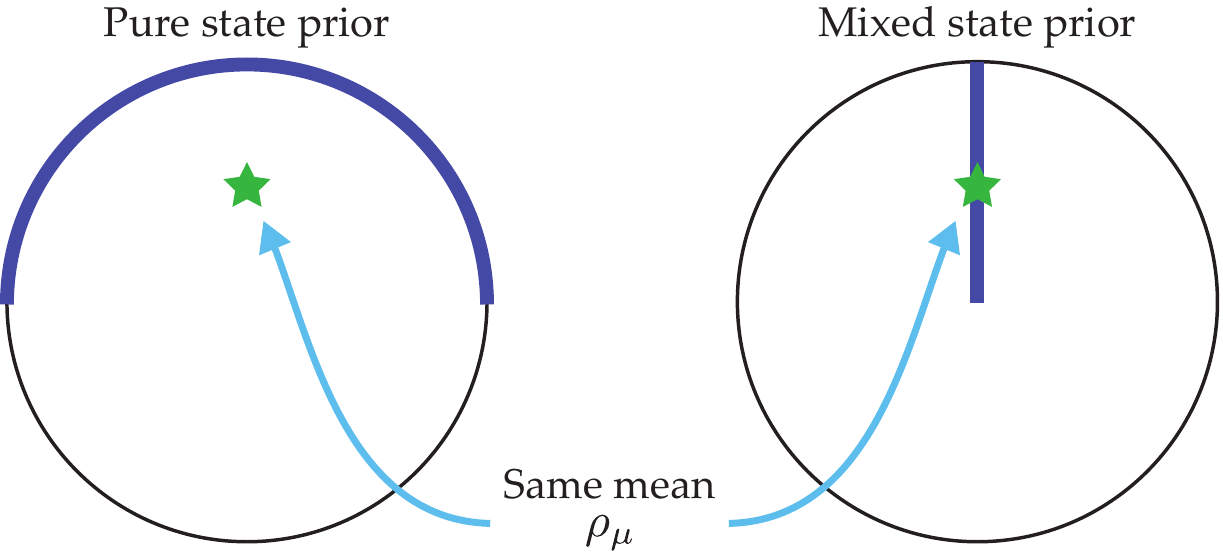}
    \end{centering}
    \caption{\label{fig:two-priors}
        Two priors, one mean state. The prior on the left assumes support only on pure states while the one on the right includes support on mixed
        preparations. We choose to illustrate our manuscript with rebits for visual clarity.
    }
\end{figure}

Here, we seek to use the default priors from \sec{uninformative-priors} to construct a prior $\pi(\rho)$ over states with that has a desired mean
 $\rho_\mu = \expect_{\rho\sim\pi}[\rho]$, and introduces little other information.
The choice of the mean could be informed by, for example, 
 experimental design or previous experimental estimates. Critically, a prior $\pi$ is \emph{not} uniquely
specified by the first moment of $\rho$ over $\pi$, $\rho_\mu =
\expect_{\rho\sim\pi}[\rho]$. Rather, the mean state $\rho_\mu$ only is a
complete specification of observables measured against a state drawn from the
prior. Indeed, different sets of assumptions can result in the same mean
state, as illustrated in \fig{two-priors}. Thus, additional constraints are required to
select an appropriate prior.

We also require that $\pi$ has support over all feasible states; for instance, those states of the appropriate
dimension \cite{shang_optimal_2013}, possibly subject to rank restrictions.
By making this demand, our tomography procedure can recover from bad priors, given
sufficient data; we will show the robustness of our algorithm in later in this section as well as in \sec{examples}.
Finally, we demand that our insightful priors can be
sampled efficiently with the dimension of the state under consideration.

\subsection{Construction of Insightful Priors}

To achieve the desiderata that all feasible states are supported, that we can
sample efficiently, and that the prior mean is $\rho_\mu$, we proceed in two
steps. First, we sample $\rho_f$ from an fiducial  prior
 $\phi(\rho)$ i.e. $\rho_f \sim \phi$. The sample
from the fiducial prior is then transformed to a sample from the insightful prior
under a generalized amplitude damping channel (GAD)
\begin{equation}\label{eq:gadfli}
    \rho_{\rm sample} = \Phi(\rho_f| \epsilon, \rho_*) = (1 - \epsilon) \rho_f + \epsilon \rho_* \Tr[\rho_f],
\end{equation}
where $\epsilon\in[0,1]$ is the damping parameter and $\rho_*$ is the fixed
point of the map. For $\epsilon =0$, the map is the identity channel, while
for $\epsilon = 1$, the map damps to the mixed state $\rho_*$. In our method
$\epsilon$ is not a fixed number but is drawn from an ensemble described by
the beta distribution, i.e.  $\epsilon \sim {\rm Beta}(\alpha, \beta)$.

Thus to determine the mean of $\pi(\rho)$ we must determine $\rho_\mu$ given  $\rho\sim\phi$ and $\epsilon \sim {\rm Beta}(\alpha, \beta)$, that is
\begin{align}
 \rho_\mu &:= \expect_{\rho,\epsilon}  [ \Phi(\rho| \epsilon, \rho_*)] \nonumber\\
  &=  \expect_{\epsilon} \left [  (1 - \epsilon)\frac \Id D + \epsilon \rho_* \right ]\, \nonumber\\
  &=   \frac{\beta}{\alpha+\beta} \frac\Id D + \frac{\alpha}{\alpha+\beta} \rho_* \,
\end{align}
where on the second line we have used the fact that $ \expect_{\rho\sim\pi}[\rho]= \Id/D$ for priors that are fiducial, and on the third line we have used $\expect_{\epsilon \sim {\rm Beta}}[\epsilon]=\alpha/(\alpha +\beta)$.  Inverting this relationship tells one how to choose the fixed point of the channel \eq{gadfli} to obtain a given mean
\begin{align}\label{eq:chan_fixed_point}
 \rho_*  = \frac{\alpha+\beta}{\alpha} \left (  \rho_\mu - \frac{\beta}{\alpha+\beta}  \frac\Id D \right).
\end{align}
Clearly we need more than the first moment $\rho_\mu$ to specify the
prior; we must determine $\alpha$ and $\beta$ to complete the specification. The first constraint on 
$\alpha$ and $\beta$  comes from the positivity of $\rho_*$, which is a valid state only if 
\begin{align}
 \frac{\alpha+\beta}{\alpha} \left (  \lambda_i - \frac{\beta}{\alpha+\beta}  \frac1 D \right)>0 \quad \forall i,
\end{align}
where $\lambda_i$ are the eigenvalues of  $\rho_\mu$. Thus, the minimum eigenvalue $\lambda_{\min}$ of $\rho_\mu$ partially constrains $\alpha$ and $\beta$ by  $\lambda_{\min}>\beta/[D(\alpha+\beta)]$. In order to completely determine the parameters of the beta distribution, we adopt the principle that the action of the channel \eq{gadfli} should be minimized.
We use this principle as an efficient heuristic motivated by analogy with maximum entropy methods.
In other words, \emph{the insightful prior is as uninformative as possible
given the constraint of the chosen mean, and with respect to a particular default prior}.

We therefore choose to
minimize the expected value $\expect_{\epsilon \sim {\rm Beta}}[\epsilon]=\alpha/(\alpha +\beta)$, such that $\pi$ is the
closest GAD-transformed distribution to $\phi$ with the given mean $\rho_\mu$. This
minimization gives 
\begin{align}\label{eq:gadfli_beta}
\alpha = 1 \quad {\rm and}\quad \beta= 
\frac{D \lambda_{\min}}{1-D \hspace{0.5pt}\lambda_{\min} }.
\end{align} 

This construction naturally specializes to provide a procedure for estimating
the bias of a coin, as discussed in \apx{coin-gadfli}. 

\subsection{Convexity and Robustness of Insightful Priors}

Note that, because
$\epsilon = 0$ is in the support of all beta distributions, our prior ensures
that its support is at least that of the
given fiducial prior, $\supp \pi \supseteq \supp \phi$.
For the default priors listed in \sec{uninformative-priors},
the prior constructed by our procedure has support over all states of the appropriate
dimension. In general, the fiducial prior \emph{defines} the
states that we consider to be valid, as can be seen from the convexity
of our construction.

That is, if $\rho_\mu$ is a convex combination over states in
the support of the fiducial prior, then $\supp \pi$ is the convex closure of $\supp \phi$.
On the other hand, if $\rho_\mu$ lies outside of the support of the fiducial prior, then our algorithm
chooses $\rho_*$ to lie outside as well, such that the support of the insightful prior is
bigger than that of the convex closure of the fiducial support.

As our procedure preserves the support of the fiducial prior in both
cases, our procedure also defines insightful priors for rebits and channels by using
the real Ginibre and BCSZ priors as fiducial priors, respectively.
In particular, using the BCSZ prior as the fiducial prior for a GAD-transformed
distribution, we then obtain a prior that is insightful and is supported over all completely-positive
and trace-preserving maps of a given dimension. Together with the \cj~isomorphism
described in \apx{choi}, we can apply SMC to \emph{process} tomography with little
further effort.

\begin{figure}
    \begin{center}
        \includegraphics[width=\columnwidth]{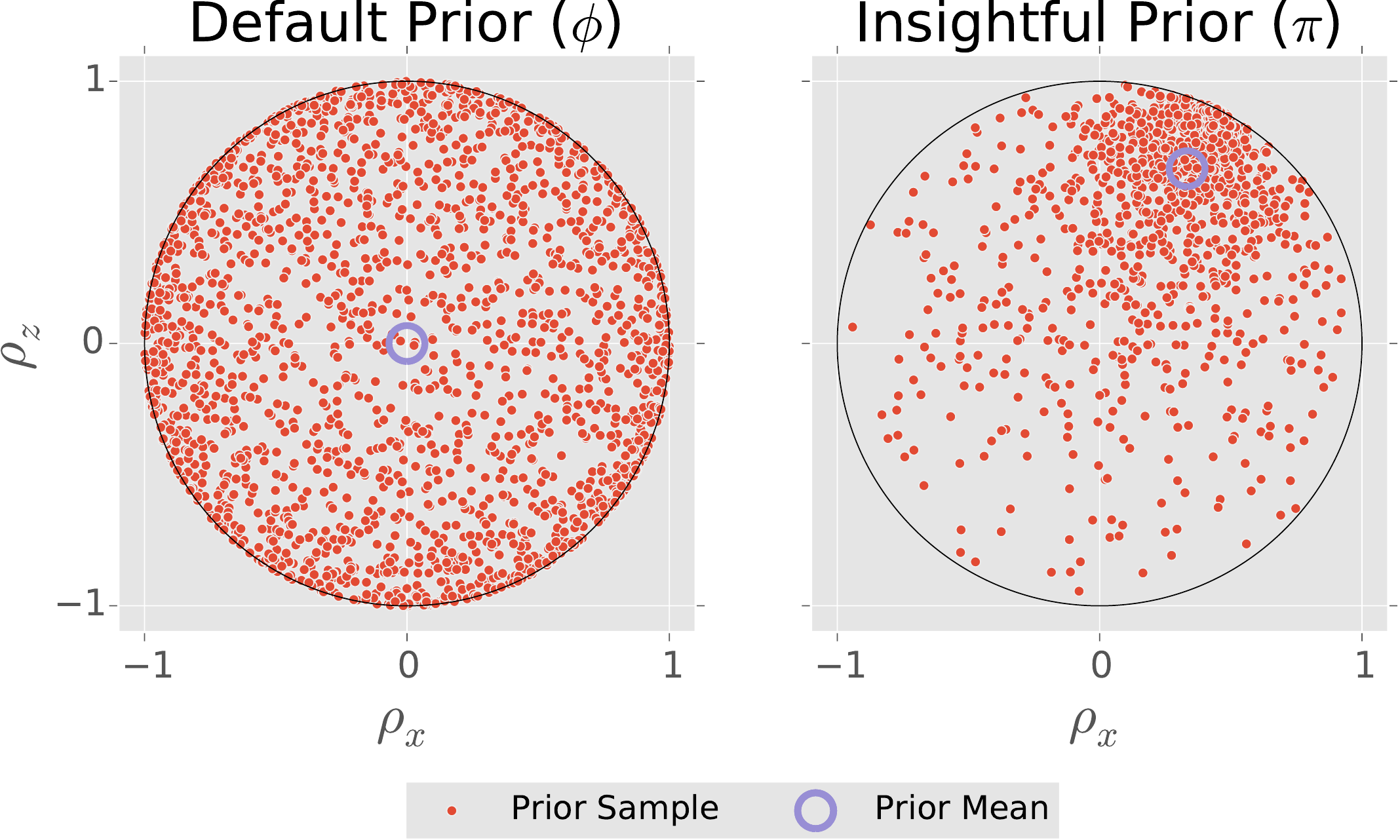}
    \end{center}
    \caption{
        \label{fig:gadfli-rebit}  
        (Left) A default prior (rank-2 Ginibre ensemble) for a 
        single rebit. This fiducial prior is 
        transformed into a prior that is insightful (right) by first choosing a mean, in this case $\rho_\mu = \frac12(\id + \frac23 \sigma_z + \frac13 \sigma_x)$.
        Next, the generalized amplitude damping channel consistent with this
        mean state, as in \eq{gadfli}, is applied to the samples. 
    }
\end{figure}

More exotic
fiducial priors, such as distributions over stabilizer states or mundane states in the sense described by Veitch \etal~(zero-mana) \cite{veitch_resource_2014}, can also
be used, provided that they can be efficiently sampled and have the maximally-mixed state
as their mean. For example, a prior that is insightful for mixed rebits is given in
\fig{gadfli-rebit}.

Before proceeding, however, we note that our notion of a fiducial
prior does not imply that such priors are \emph{uninformative}---
indeed, we have seen that they serve to define what states are considered valid
at all.
Indeed, as Wasserman states \cite{Wasserman_priors},
\emph{``by definition, a prior represents information. So
it should come as no surprise that a prior cannot represent lack of
information.''}
As a further example, consider a prior over states of a given purity $r$; for a qubit, such
states form a shell inside of the Bloch sphere.
This prior conveys
conveys information
about what states are considered feasible at all, but still reports as its initial
estimate that the state of interest is maximally mixed, i.e. it is fiducial, and can
be used as input to our algorithm. 
Indeed, were one to do so, our algorithm would use this specification of what a feasible state is to
define an insightful prior that is limited to a convex hull of the ball of states of purity no greater
than $r$ and the desired mean state (provided $\Tr(\rho_\mu^2)\le r$, such
that $\rho_\mu$ lies within the given purity ball).
Taking the case as $r \to 1/D$ (that is, a $\delta$-function
prior supported only at the maximally-mixed state), the situation becomes more extreme, in
that the insightful prior is then supported only on the line connecting the maximally-mixed state
to $\rho_*$.
For this reason, the default priors given in \sec{uninformative-priors}
are chosen to have support
over all states of a given dimension and rank, making them
especially useful inputs to our algorithm.

\begin{figure}
    \begin{center}
        \includegraphics[width=\columnwidth]{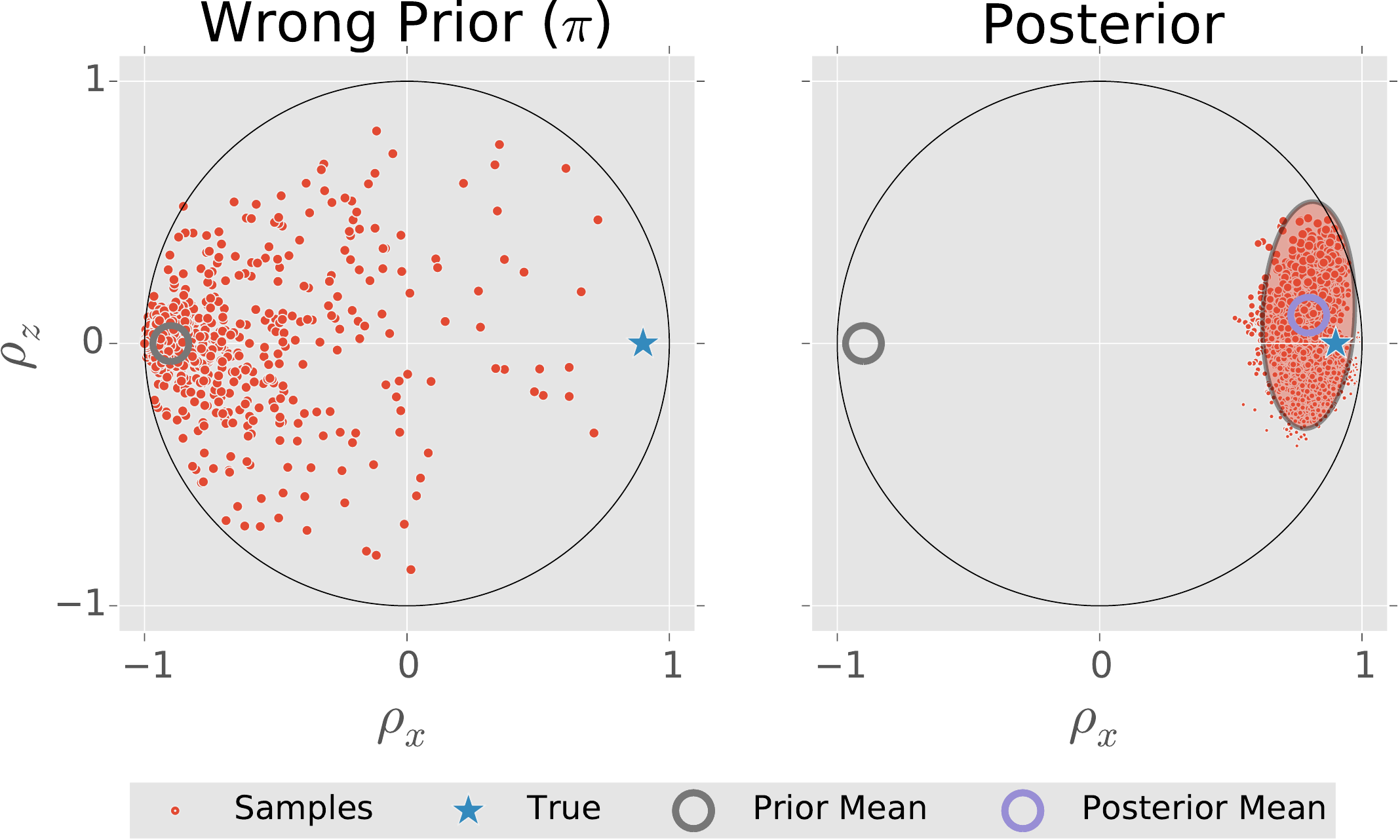}
    \end{center}
    \caption{
        \label{fig:wrong-prior}  
        A demonstration of the robustness of our insightful prior. 
        (Left) A prior that is insightful in the mean $\rho_\mu = \frac12(\id - \frac{9}{10} \sigma_x)$. 
        The true state, $\rho_{\rm true}=\frac12(\id + \frac{9}{10} \sigma_x)$, is almost orthogonal to the mean of the prior
        and has purity $\Tr[\rho_{\rm true}^2]=0.905$.
        (Right) After 30 random Pauli measurements (10 shots each) the posterior mean is centered on the true state.
        A covariance ellipse at three standard deviations is drawn in as well, indicating the normal approximation
        to a $99\%$ credible region. Notice that the ellipse slightly leaks out of the state space; the SMC approximation
        is rich enough to avoid this problem.
        An animated version of this figure is available online \cite{granade_quantum_2015b}.
    }
\end{figure}

Finally, it is vital that the prior we have suggested is robust. In \fig
{wrong-prior} we choose the mean of  our insightful prior to be almost
orthogonal to the true state. After 300 random Pauli measurements, the
posterior has support on the true state and the mean of the posterior is
approximately the true state. Thus, even if the mean of the prior is woefully
wrong our procedure is robust in that it provides a reliable estimate. This
robustness to a bad initial prior requires additional data to be collected,
such that useful prior information can accelerate experiments but will not, in
general, lead to wrong conclusions. We explore this robustness further in
\sec{examples}.

\section{Tomographic State Tracking}
\label{sec:state_space_tracking}

In this section we use a generalization of particle filtering methods to 
track a stochastic processes. In the context of quantum state tomography,
{\em state-space methods} allow us
to characterize a stochastically-evolving source without having to ignore
all previous data. We call the resulting method \emph{tomographic state
tracking}.

In particular we use the \textsc{condensation} algorithm, 
which interlaces Bayes updates with updates to move sequential Monte Carlo particles 
(using drift and diffusing of the particles), to follow a stochastic process
\cite{IsaBla98}. This technique has since been applied in a variety of other
classical contexts \cite{doucet_sequential_2000,jasra_sequential_2009} as well as in quantum
information \cite{granade_characterization_2015}. Such methods, collectively
known as \emph{state-space} particle filtering, are useful for following the
evolution of a stochastic process observed through a noisy measurement.

\begin{figure}
    \begin{center}
        \includegraphics[width=0.95\columnwidth]{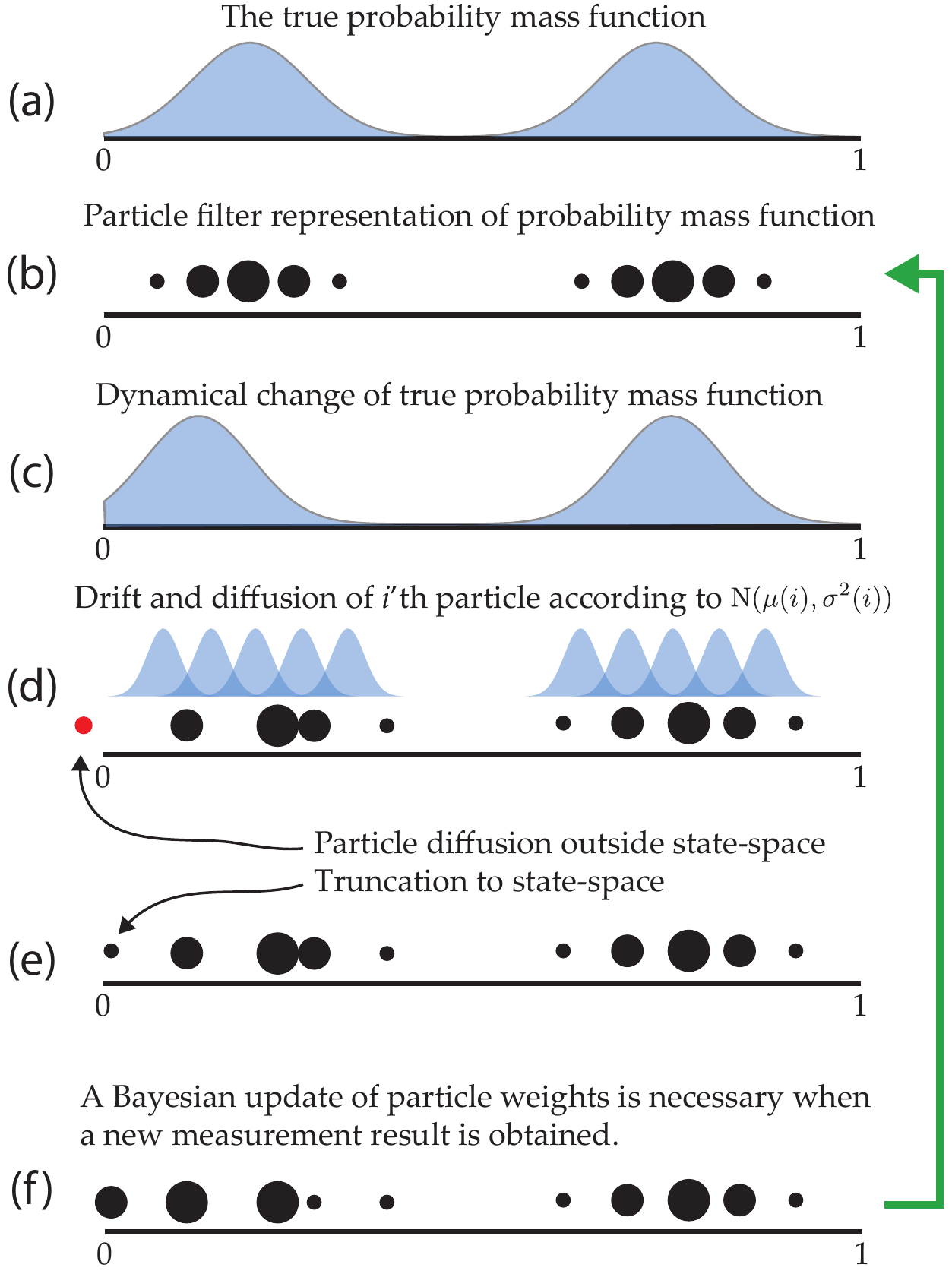}
    \end{center}
    \caption{\label{fig:coin-state-particle-tracking} 
        Illustration of state-space tracking for particle filters on
        a coin state via  the \textsc{condensation} algorithm~\cite{IsaBla98}. 
        From top to bottom, (a) we start with a continuous posterior
        over state space, (b) then discretize by sampling to obtain our initial SMC /
        particle filter approximation, (c) a dynamical change of the true distribution occurs,
        (d) we then perturb each particle by a Gaussian
        and (e) truncate to valid probabilities to complete the diffusion
        step, and (f) perform a Bayes update on the next datum. The final
        posterior approximation then forms the new approximation (b) for
        the next diffusive update. As explained in the main text, the Bayes update
        actually updates the joint distribution of the parameters to be estimated and the drift
        and diffusion parameters. This means the drift and diffusion ``co-evolve'' with the posterior, 
        which is at the heart of the  \textsc{condensation} algorithm.
    }
\end{figure}

\begin{figure*}
    \begin{center}
        \includegraphics[width=\textwidth]{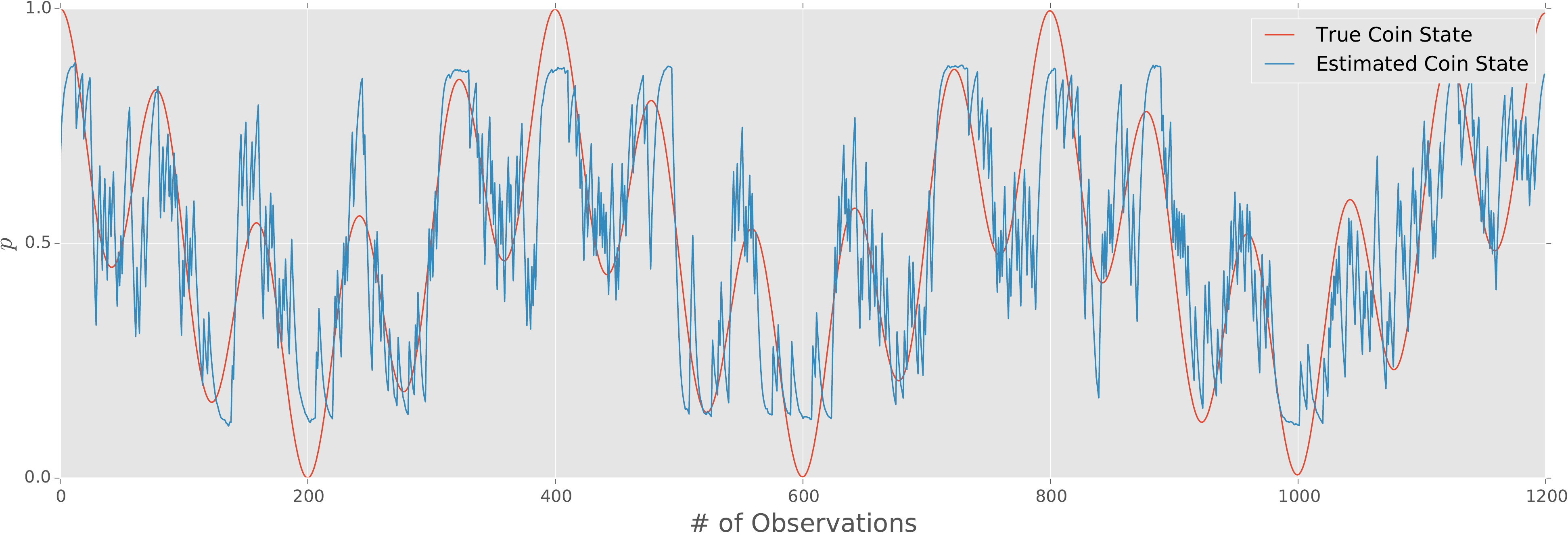}
    \end{center}
    \caption{
        \label{fig:coin-diffusion}
        Diffusive tracking of a coin state (bias). This figure illustrates that the \textsc{condensation} 
        algorithm can track a deterministic process even when ``drift'' terms in the \textsc{condensation} 
        algorithm are not updated. Here, we consider a coin with the true coin state evolving as a 
        two-tone sinusoidal function  $p(t_k) = \frac14[2 + \cos(2 \pi   f_1  t_k) + \cos(2  \pi f_2 t_k)]$,
        sampled at discrete times $t_k= k \Delta t$ where $\Delta t$ is 
        the time between samples.  In this figure, the frequencies are $f_1= 1/80$ and $f_2= 1/294$.
    }
\end{figure*}

We now briefly explain the \textsc{condensation} algorithm, readers interested in further
details are directed to the original paper \cite{IsaBla98}. Consider \fig{coin-state-particle-tracking} which illustrates
the \textsc{condensation} algorithm for tracking a coin with a dynamical bias $\Pr({\rm Heads})=p(t)$. The current posterior of the coin bias, i.e. \fig{coin-state-particle-tracking}(a), is given an SMC representation in \fig{coin-state-particle-tracking}(b). In \fig{coin-state-particle-tracking}(c), the true probability mass function changes---
at this point, we step through the \textsc{condensation} algorithm to track this change.
Each SMC particle is 
perturbed by a Gaussian. In particular, the $i$th particle is perturbed by a Gaussian with mean 
$\mu(i)$ and variance $\sigma^2(i)$.

In keeping with the terminology used by Isard and Blake \cite{IsaBla98},
the mean of the perturbation is termed the \emph{drift}, and allows one to track 
deterministic evolution of a probability mass function; this becomes evident if $\sigma^2(i)=0$ for all $i$.
Similarly, the variance of the perturbation is termed \emph{diffusion} and 
will allow the algorithm to track a stochastic process.
As we will detail further
below, both the drift and diffusion parameters can be learned online, such
that we do not require them to be known \emph{a priori}.

Sometimes the perturbation by the Gaussian will cause the 
particles to fall outside of the state space, in this case the unit interval, see e.g. \fig{coin-state-particle-tracking}(d). In this situation, we modify the 
\textsc{condensation} algorithm to truncate particles to be valid probabilities,
completing the Gaussian perturbation step, see \fig{coin-state-particle-tracking}(e). Finally, 
we obtain the next datum and perform a Bayes update on the next datum. The final posterior approximation, 
 \fig{coin-state-particle-tracking}(f), then forms the new approximation (b) for the next diffusive update.

 Interestingly, the \textsc{condensation} 
 algorithm starts with a joint distribution over the parameters to be estimated.
 For the coin case, these are the bias $p(t)$, and the distribution parameters
 for drift and diffusion
 $\Nor(\mu(i),\sigma^2(i))$. Thus, when the Bayes update occurs the drift and diffusion parameters are updated as well, 
 even though the likelihood does not explicitly depend on these parameters, which is referred to as co-evolution. It is this co-evolution that 
 powers the tracking capabilities of the \textsc{condensation} algorithm.

As the learning of deterministic evolution of states
is well-understood
\cite{schirmer_quantum_2010,sergeevich_characterization_2011,granade_robust_2012},
we will suppose that the state
under study is with respect to a frame that has already been well-characterized.
Thus, the dominant remaining dynamics of the state under study are stochastic,
such that we need not consider drift updates in our state-space model. 
Even with this assumption our model can still track deterministic evolution, however,
provided that the diffusion is strong enough to include the true evolution with
high probability (see \fig{coin-diffusion} for an example of tracking deterministic
evolution with diffusion alone).

Concretely, we update particle $i$ by $\rho_i(t_k) \mapsto \rho_i(t_{k+1})$ by first adding drift
and diffusion terms to find a step $\Delta\rho_i(t_k)$ to take in state-space,
then truncating the negative eigenvalues of $\rho_i(t_k) + \Delta\rho_i(t_k)$. For
each particle $\rho_i$, we let $\Delta\rho_i(t_k) = \Delta\mu + \Delta\eta$,
where $\Delta\mu = \Delta\mu(t_k)$ is a deterministic drift, and where each
traceless component of $\Delta\eta$ is drawn from a Gaussian $\Nor(0,
\sigma^2)$ with standard deviation $\sigma$. As stated above, we work in a
frame where the deterministic part has been taken out so that $\Delta \mu = 0$
for all particles and for all time.
The diffusion standard deviation $\sigma$ is
then taken to be a function of evolution time and the new model parameter $\eta$,
$\sigma = \sqrt{t_{k+1} - t_k} \eta$. This allows the evolution rate to ``co-evolve''
with the state model $\rho$, as described above.

Diffusion is completed by finding the spectral decomposition of $\rho_i(t_k) +
\Delta\rho_i(t_k)$, then
truncating and renormalizing. In particular, let $\rho_i(t_k) + \Delta\rho_i(t_k)
= \sum_{j} \lambda_{i,j} \ket{\psi_{i,j}}\bra{\psi_{i,j}}$. Then,
\begin{equation}
    \rho_i(t_{k+1}) = \frac{1}{\mathcal{N}_i} \sum_{j} \begin{cases}
        \lambda_{i,j} \ket{\psi_{i,j}}\bra{\psi_{i,j}} & \text{if } \lambda_{i,j} \ge 0 \\
        0                                              & \text{if } \lambda_{i,j} <   0
    \end{cases},
\end{equation}
where $\mathcal{N}_i$ is chosen such that $\Tr(\rho_i(t_{k+1})) = 1$.
This truncation rule avoids expensive optimization to find the closest
state consistent with a given drift and diffusion update $\Delta\rho_i(t_k)$,
while still generalizing methods known to be effective and efficient for
coin estimation.

In order to determine the limitations of state space tracking,  we considered
tracking a single tone cosine $p(t_k) = \cos(2\pi f k \Delta t)$, where $t_k =
k \Delta t$ is a discrete time and $\Delta t$ is the time between  samples.
Recall that we are choosing not to include deterministic evolution (i.e.
``drift'') in our model, thus the following observation only applies to purely
diffusive tracking. We found the our algorithm could track a frequency up to
$f_{\rm track}=(1/10)\times(1/\Delta t)$. At higher frequencies, our approach
failed to track the oscillatory behavior of $p(t)$, in that it would report
$p(t_k) = 1/2$ for all time. This modality can be tested using model selection
\cite{enk_when_2013,granade_characterization_2015}, such that more a
appropriate algorithm can be used in that case. A more  sophisticated, though
less quantitative, analysis of this failure can be found in \apx{track_freq}.

\section{Numerical Examples}
\label{sec:examples}

We now show examples of Bayesian state and process tomography using sequential
Monte Carlo, with priors that are respectively default and insightful.
These examples were generated using QInfer \cite{granade_qinfer:_2012}.

\begin{figure}
    \begin{center}
        \includegraphics[width=\columnwidth]{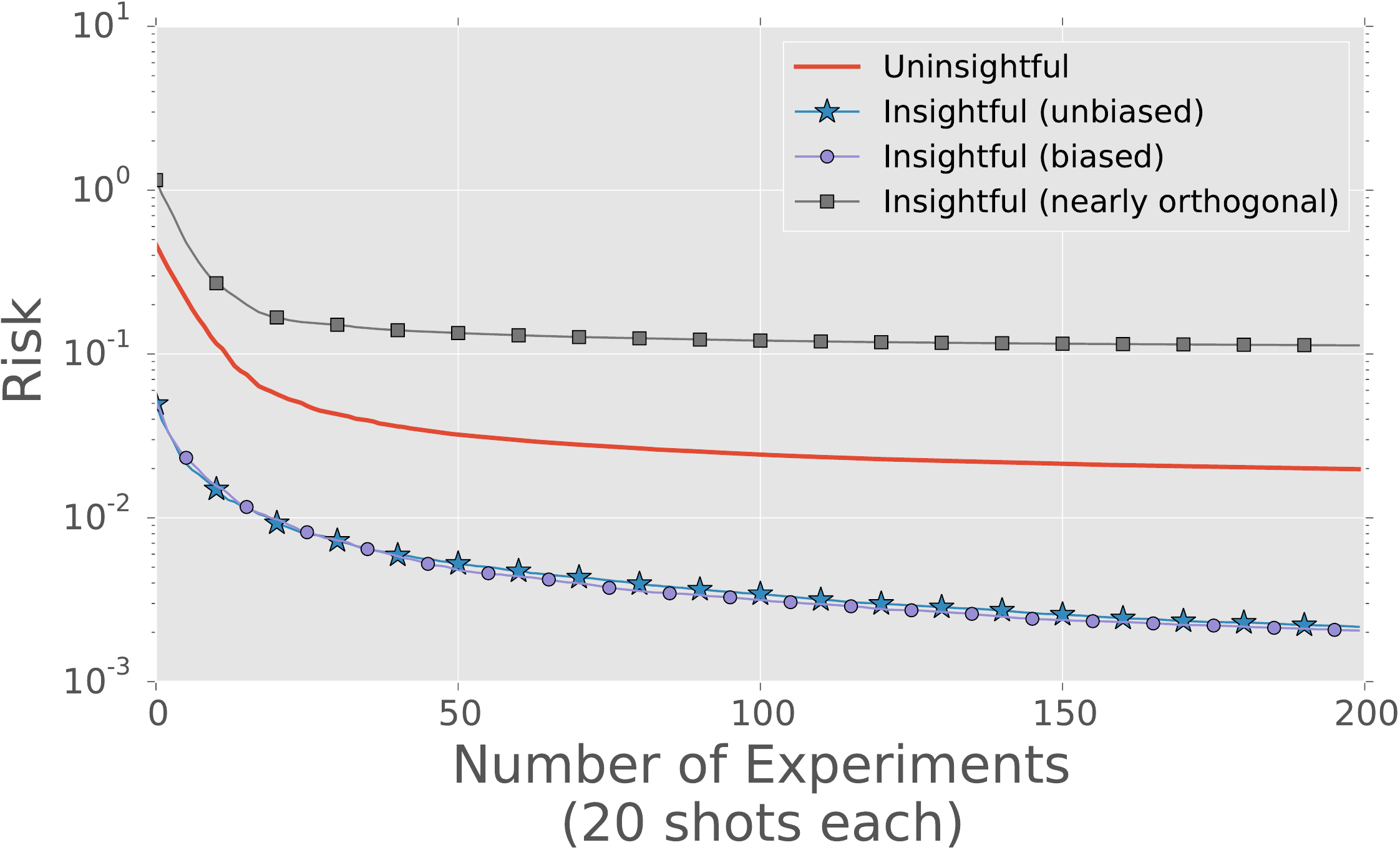}
    \end{center}
    \caption{
        \label{fig:qutrit-risk}
            Risk of SMC-based tomography for a qutrit, using three different priors,
            with true states drawn from the generalized amplitude damping distribution
            for $\rho_\mu = \diag(0.9, 0.05, 0.05)$.
            The default prior is taken to be the full-rank qutrit Ginibre distribution,
            while the insightful prior is taken to be identical to the actual distribution
            over true states and the biased prior uses the mean $\rho_\mu = \diag(0.87, 0.065, 0.065)$.
            The nearly orthogonal prior is taken to be insightful with the mean
            $\rho_\mu = \diag(0.065, 0.065, 0.87)$.
            Risk is measured with respect to the quadratic loss function on
            vectorized density operators, $L(\hat{\rho}, \rho) = \|\sket{\hat{\rho}} - \sket{\rho}\|_2$.
            The risk is averaged over 1200 trials. Measurements are chosen to be rank-1 projectors
            onto randomly drawn single-qutrit stabilizer states. Each such stabilizer state is then measured
            20 times.
    }
\end{figure}

For state tomography, we demonstrate our methods by learning qutrit states,
as shown in \fig{qutrit-risk}. We demonstrate the performance of our algorithm
in this case by reporting the \emph{risk}, defined as the expected quadratic
loss over repetitions of the algorithm,
\begin{equation}
    \label{eq:risk}
    r(\hat{\rho}, \pi) = \expect_{\rho\sim\pi} [\|\sket{\hat{\rho}} - \sket{\rho}\|_2].
\end{equation}
We use each of a default, insightful and unbiased, and an insightful
but biased prior. In all three cases, we draw the ``true'' states $\rho$ from
the prior that matches the insightful and unbiased prior.

In \fig{qutrit-risk}, we also verify that our method is robust for the qutrit
example by considering an insightful prior whose mean is nearly orthogonal to
the true state. Notably, in well over half of the 1200 trials considered in that case,
QInfer reported that the algorithm was likely to fail, heralding the impact of the
``bad'' prior.

We then proceed to consider state-space quantum state tomography, as detailed
in \sec{state_space_tracking}.
We demonstrate the performance of our state-space method
in an animation, available online \cite{granade_quantum_2015}, also see \vid{diffusive-tracking} for a snapshot.

\begin{figure*}
    \begin{center}
        \includegraphics[width=\textwidth]{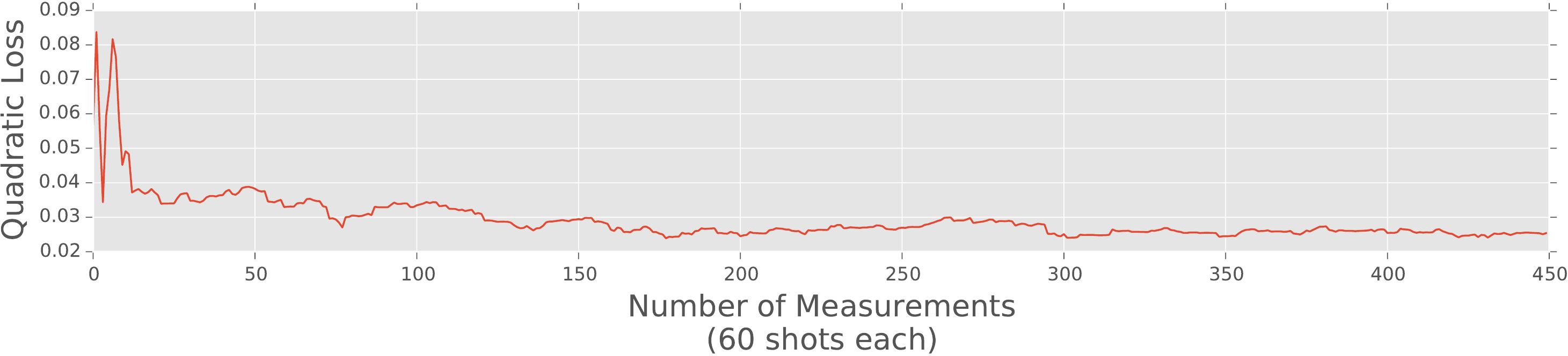}
    \end{center}
    \caption{
        \label{fig:qpt-loss}        
        Example quadratic loss between estimated and true channels for SMC-based tomography
        for a qubit channel $\Lambda[\rho] = 0.7 \rho + 0.3 H\rho H$, where $H$ is
        a Hadamard. The prior for both the SMC
        estimator and the true channel is taken to be the insightful prior
        with a mean taken to be a 90\% mixture of the true channel and 10\% of
        a completely depolarizing channel. Out of the 450 experiments,
        20\% are chosen to be random Pauli preparation and measurements,
        while
        80\% are chosen adaptively. In particular, adaptation proceeds
        by proposing 50
        different random pairs of Pauli preparations and measurements, then
        selecting the one which has the maximum expectation value
        $\sbraket{\Sigma} = \sbraket{P, M | \Sigma | P, M}$, where $\Sigma$ is
        the current covariance superoperator.
    }
\end{figure*}

\begin{figure*}
    \begin{center}
        \includegraphics[width=0.9\textwidth]{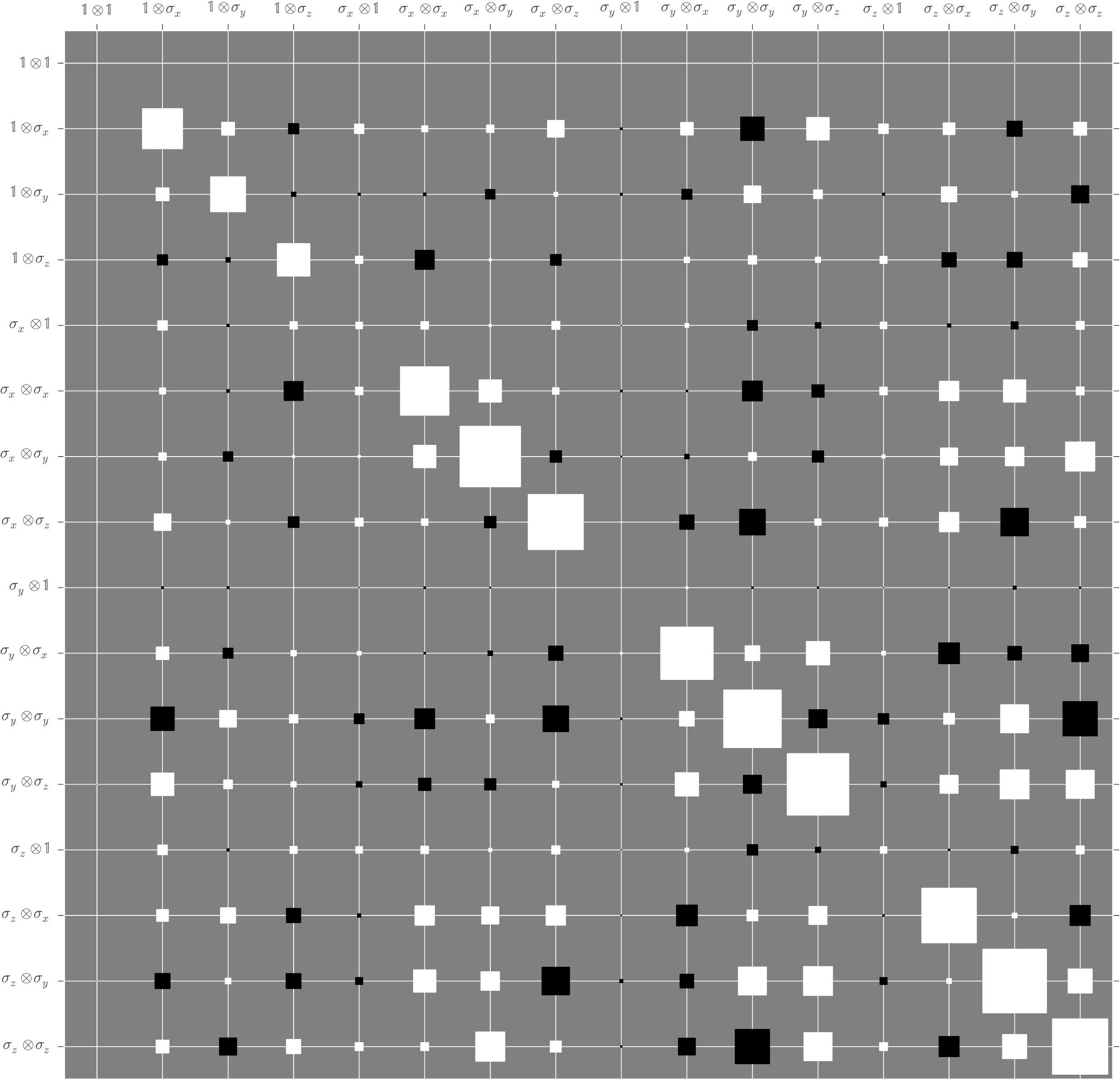}
    \mbox{
        \includegraphics[width=0.3\textwidth]{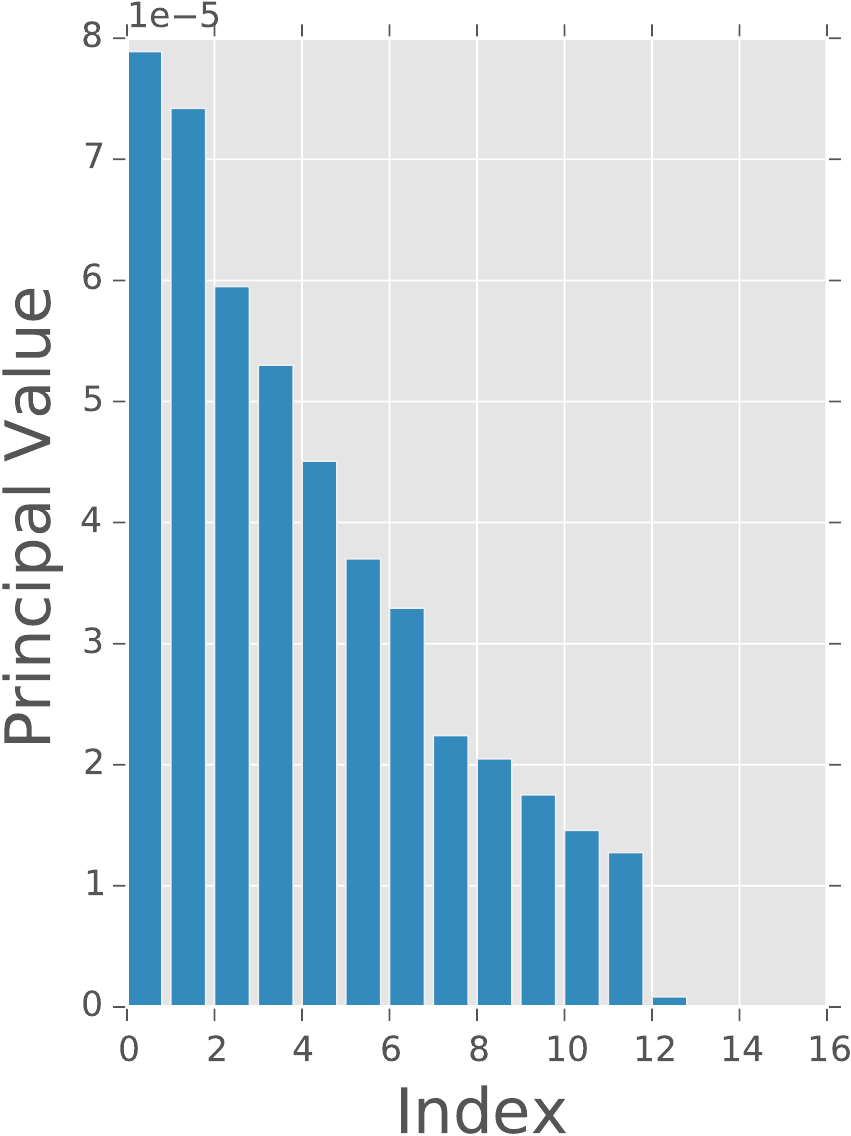}       
        \hspace{10mm}
        \includegraphics[width=0.5\textwidth]{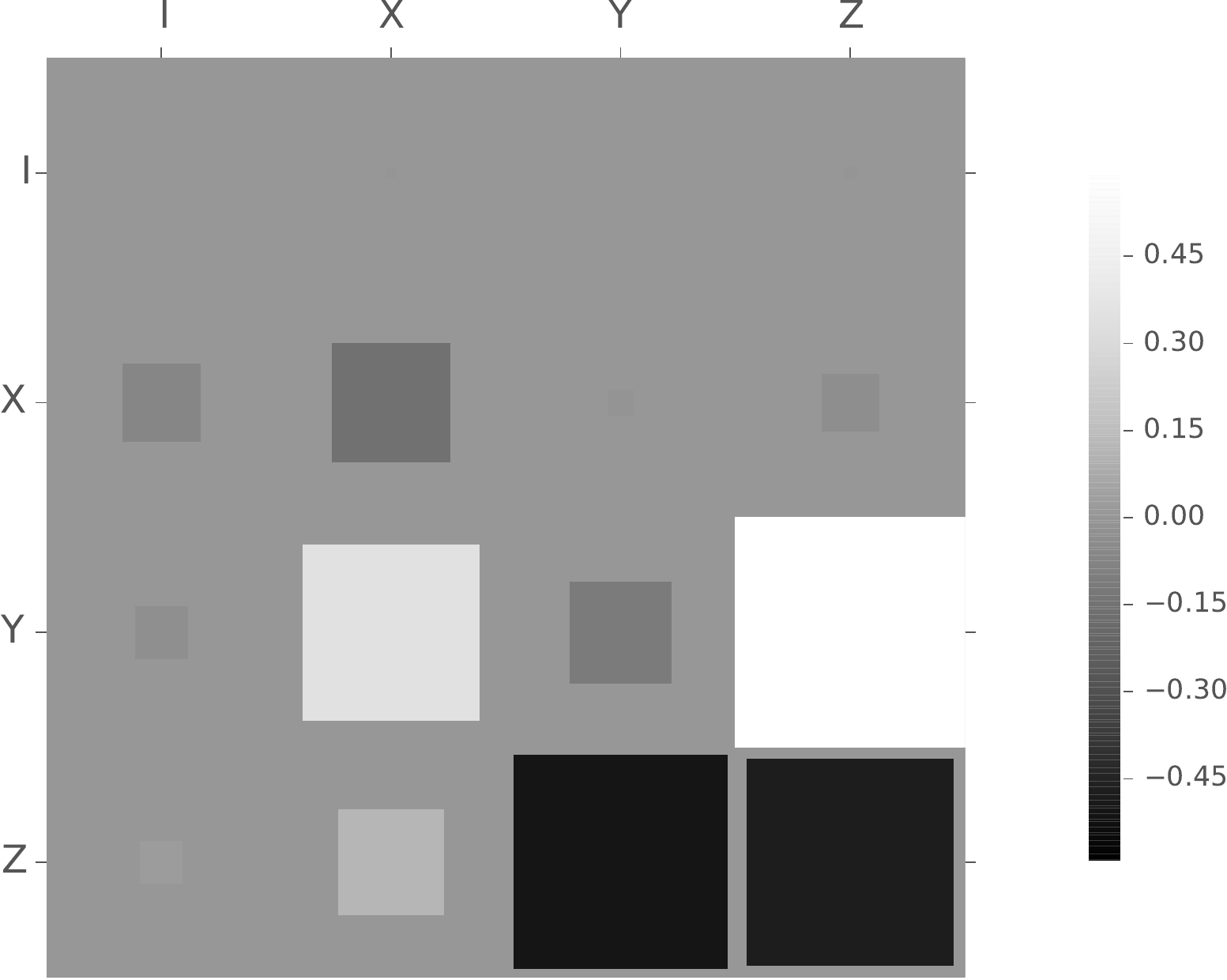}
        }
    \end{center}
    \caption{
        \label{fig:qpt-cov}
        (Top) Covariance matrix for process tomography experiment shown in \fig{qpt-loss}.
        (Bottom left) Spectrum of covariance matrix (principal values).
        (Bottom right) Largest eigenchannel of the covariance matrix (principal channel), indicating
            the direction in which the channel is least certain.
    }
\end{figure*}

Finally, we demonstrate the application of our method to learning quantum
channels acting on a qubit. In \fig{qpt-loss}, we show an example of
a single simulated quantum process tomography experiment, where the true channel and the insightful
prior (generated with a BCSZ fiducial prior) agree. The preparation and
measurement settings are chosen to be elements of the Pauli basis. 
Specifically, 20\% of the experiments use random Pauli preparation and measurements, while 80\% of the experiment use settings that maximize
$\sbraket{\rho_i^\T, E_i | \Sigma \rho | \rho^\T_i, E_i}$ out of 50 randomly proposed Pauli preparations
and measurements; that is, adaptively chosen to overlap with the principal components
of the current posterior.
The resulting posterior distribution characterizes the uncertainty remaining
about the ``true'' channel, as shown in \fig{qpt-cov}. In particular, we note
the principal components of the posterior covariance matrix are themselves
quantum maps which describe the directions of maximal uncertainty in the final
posterior. For this example, our error is dominated by uncertainty about the
contribution of the identity and Hadamard channels, as is made clear by visual inspection
in \fig{qpt-cov} (bottom right).

\begin{video*}
    \begin{center}
        \includegraphics[width=\textwidth]{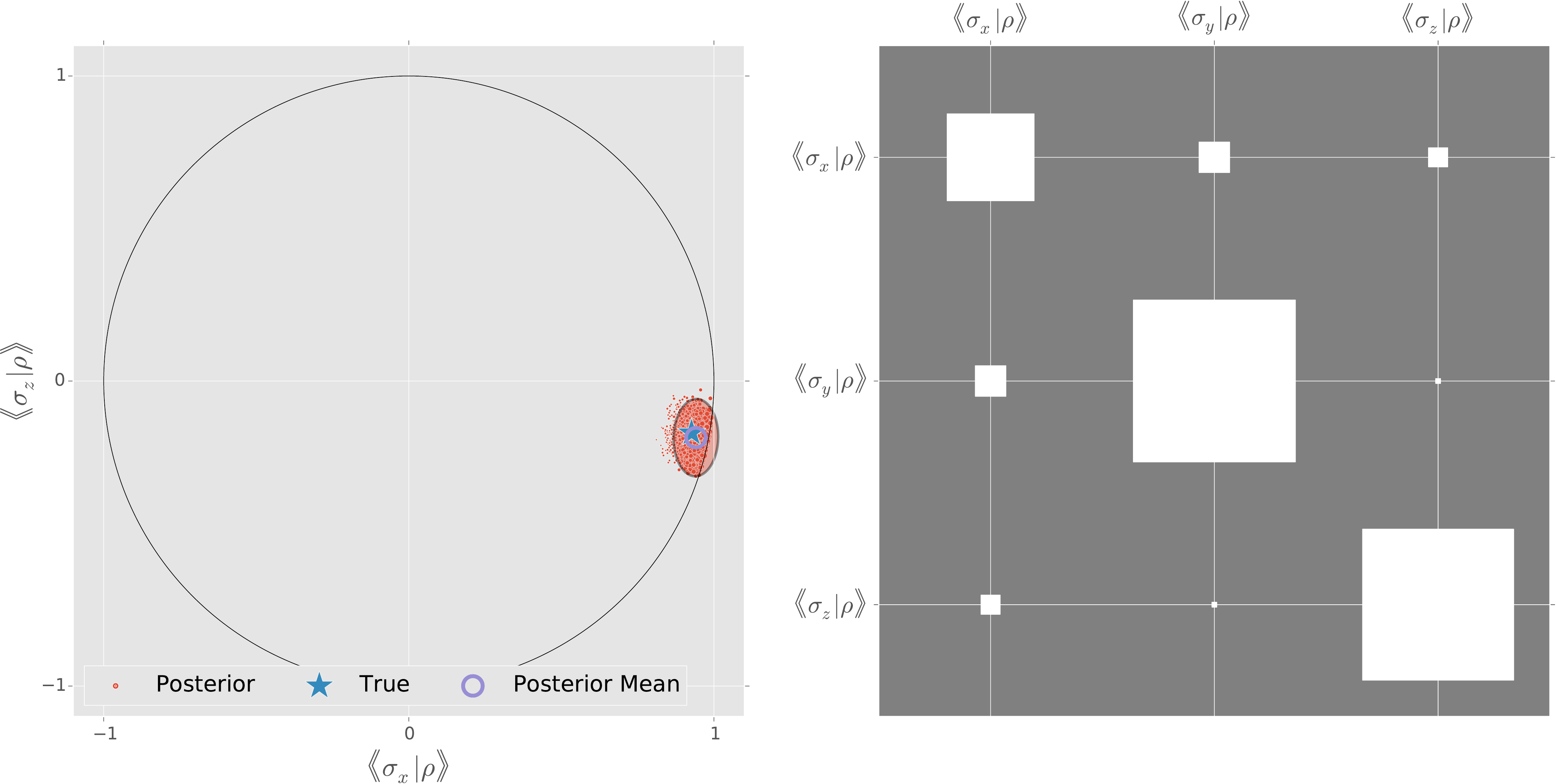}
    \end{center}
    \setfloatlink{https://goo.gl/mkibti}
    \caption{\label{vid:diffusive-tracking}
This video demonstrates using tomographic state tracking to estimate a diffusing quantum state. Shown at left is the projection of the true state (star), the sequential Monte Carlo particle cloud (red dots), and covariance approximation to the 99\% credible region (red circle) to the $x$-$z$ plane. Shown at right is a Hinton diagram of the current posterior covariance matrix.
        At each time step, each element of the vectorized true state is perturbed by a Gaussian with mean 0 and standard deviation 0.0045, and then is truncated to lie within the space of valid quantum states. The prior over this diffusion rate is taken to be a log-normal distribution with mean 0.006. Each measurement consists of 25 shots along a random Pauli axis.
    }
\end{video*}

\section{Conclusions}
\label{sec:conclusion}

In this work, we have provided a new prior distribution over quantum
states and channels that allows for including experimental insight, a software
implementation for numerically approximating Bayesian tomography, and a method
for tracking time-dependent states. Together, our advances make Bayesian
quantum tomography practical for current and future experiments. In
particular, our methods allow for exploiting well-known benefits of Bayesian
methods, including credible region estimation, hyperparameterization and model
selection.

We note, however, that our insightful prior on states and
channels is completely specified by its first moment. An interesting and open
problem would thus be to develop a prior on states and channels that is
completely specified by its first and second moments.

Finally, with respect to the tomographic state tracking methods presented in
\sec{state_space_tracking}, it is  worth noting that van Enk and Blume-Kohout
\cite{enk_when_2013} suggested that  model selection could be used to
determine if a source was drifting or diffusing. In this manuscript we have
provided a way that allows one to track a source that is drifting or
diffusing. It is also  possible to combine the approaches and use model
selection to determine when tracking is necessary or when a static model is
sufficient as demonstrated by Granade \cite{granade_characterization_2015}.

In short, with constructive methods for sampling from insightful priors, and
with modern statistical methods, Bayesian state and process tomography are
made practical for current experimental needs. This in turn allows us to
explore new questions in tomography, and thus better characterize and diagnose
quantum information processing systems.

\acknowledgements{
    The authors acknowledge Sarah Kaiser, Chris Ferrie, Brendon Higgins, and
    Richard K\"ung for discussions.
    The authors acknowledge funding from Industry Canada, CERC, NSERC, the
    Province of Ontario, and the CFI is gratefully acknowledged. 
    CG acknowledges Connor Jarvis for noticing an important typographical error.
    CG also acknowledges Ian Hincks, Nathan Wiebe, and Yuval Sanders
    for contributions to and testing of QInfer, as well as support from the US
    Army Research Office via grant numbers W911NF-14-1-0098 and W911NF-14-1-010.

    The authors would like to note independent related work by
    Faist and Renner \cite{faist_practical_2015}
    and by Struchalin \etal~\cite{struchalin_experimental_2015}.
}


\bibliography{pbt}


\onecolumngrid
\appendix

\newpage

\section{\cj\ Isomorphism and QPT}
\label{apx:choi}

In order to interpret \eq{likelihood1} as specifying probabilities for quantum
process tomography \cite{ChuNie97} as well as state tomography, we use the \cj\ isomorphism
and
represent a hypothesis about the channel $\Lambda$ as a state
\begin{equation}
    J(\Lambda) \defeq (\id \otimes \Lambda) (\sket{\id} \sbra{\id}).
\end{equation}
The formalism of using the \cj~isomorphism \cite{choi_completely_1975,jamiolkowski_linear_1972}
to describe open quantum processes
has also been recently detailed in
detail by Wood \etal~\cite{wood_tensor_2015} using tensor network diagrams
with particular regard to quantum tomography and its generalizations
\cite{wood_thesis_2015,ringbauer_characterizing_2014}.

In the case that quantum process tomography is carried out by ancilla-assisted
process tomography \cite{altepeter_ancilla-assisted_2003}, we are done, as $J(\Lambda)$ is, up to normalization,
the state used in AAPT. On the other hand, we can also use that for any linear
operator $X$, $\Lambda(X) = \Tr_1[J(\Lambda) (\id \otimes X^\T)]$ to represent
preparation followed by evolution under $\Lambda$ and measurement as a measurement
of $J(\Lambda) / D$ \cite{watrous_cs_2013}. In particular, suppose that a QPT experiment is performed in
which each observation consists of a measurement effect $E$ and a preparation $\rho$,
\begin{subequations}
\begin{align}
\Pr(E' | J(\Lambda)) 
    &= \Tr[E \Lambda(\rho)]\\
    &= \Tr[(\id \otimes E) J(\Lambda) (\rho^\T \otimes \id)]\\
    &= \sip{\rho^\T, E}{J(\Lambda)}\\
    &= \sip{P, E}{J(\Lambda) / D}, \label{eq:choi-sip}
\end{align}
\end{subequations}
where $P \defeq \rho^\T \cdot D$ and $E' = P \otimes E$. This now has the form of a measurement
$P \otimes E$ being made on a state $J(\Lambda) / D$, such that \eq{likelihood1}
completely describes the outcomes of both in-place and ancilla-assisted QPT experiments.
Though we used the column-stacking basis to derive this equivalence, implicitly
defining a convention for the Choi state, we can insert a unitary operator on the space
of vectorized operators to observe that \eq{choi-sip} is not dependent on the choice of basis.

The likelihood after $N$ measurements is 
\begin{align}
\mathcal L\big(J(\Lambda)\big) = \Pr\big(\mathcal M |J(\Lambda)\big)= \prod _{i=1} ^N\Pr(E'_i | J(\Lambda)),  \label{eq:like_chan}
\end{align}
where $\mathcal M$ is the string of measurement results and $E'_i$ is the observation of the $i$'th generalized measurement. Note that \eq{like_chan} is of the same form as \eq{likelihood1}, such that quantum process tomography can be treated as a special case of quantum state tomography, provided that we
use an appropriate fiducial prior.

\section{Review of Sequential Monte Carlo}
\label{apx:SMC}

The purpose of this appendix is to provide a minimal summary and point the interested reader to the relevent literature on the particle filtering and sequential Monte Carlo (SMC) algorithms as
applied to Bayesian inference. There is a wealth of information in the references that need not be reproduced here.

The basic idea of SMC is to approximate a distribution over some parameters
$\mathbf{x}$ conditioned on the $j$th data $d_j$, i.e. $\Pr(\mathbf{x}|d_j)$.
The approximation uses a weighted sum of delta-functions, specifically
\begin{align}
    \Pr(\mathbf{x}|d_j) \approx \sum_{k=1}^N w_k(d_j) \delta(\mathbf{x} - \mathbf{x}_k), \label{eq:SMC_gen}
\end{align}
where the $N$ objects $\mathbf{x}_j$ are called \emph{particles} and
$w_k(d_j)$ is the {\em weight} of the $k$th particle. Here \eq{SMC_gen} should be compared
to \eq{SMC_prior} of the main text; where we suppressed many technical and notational details. As the number of
particles $N$ increases the quality of the approximation improves. The initial
particles are chosen by sampling the prior distribution over $\mathbf{x}$,
and are taken to have uniform weight.

The next step is to update the posterior distribution given new datum say the
$j+1$th data is  $d_{j+1}$. Then using the likelihood function the particle
weights are updated from  the previous weights ($w_k(d_j)$) using
\begin{align}
    w_k(d_{j+1}) \propto \Pr(d_{j+1}|\mathbf{x}_k)w_k(d_j).
\end{align}
Clearly, the particle weights after this update need to renormalized
such that $\sum_k w_k(d_{j+1}) = 1$.

As updates proceed, the concentration of weights on the most plausible
particles causes the distribution to be samples by a much smaller number of
\emph{effective} particles, as measured by the effective sample size
$n_{\text{ess}} = 1 / \sum_k w_k^2(d_{j+1})$. Thus, periodically particle
locations have to be perturbed and the weights reset to uniform, restoring
numerical stability. This is
called resampling. A video demonstrating the Bayesian update and resampling
steps for a simple Hamiltonian learning problem is available online
\cite{granade_robust_2015}.

In the remainder of this appendix we provide references which can be used to obtained more details on SMC. Doucet \etal~have provided a more through
review from the perspective of classical statistics \cite{DouJoh09a,doucet_tutorial_2011}. In the quantum domain summaries of the SMC algorithm can be found in references \cite{granade_robust_2012,ferrie_high_2014,stenberg_efficient_2014,wiebe_hamiltonian_2014,wiebe_quantum_2014,wiebe_quantum_2015}.
Finally, Svensson has provided a useful video tutorial illustrating the application
of particle filters in a simplified radar application \cite{svensson_particle_2013}.

For our application (quantum tomography), Husz\'ar and Houlsby \cite{HH12} and
by Ferrie \cite{ferrie_quantum_2014} suggested SMC as numerical tool for
implementing \eq{rho_est}.  Of particular interest is Ferrie's recent work on
tomographic region estimators with SMC \cite{ferrie_high_2014}. Ref.
\cite{granade_characterization_2015} has provided a summary of recent
applications in quantum information.

\section{Sampling Default Priors}
\label{apx:sampling-algorithms}

In this Appendix we review the algorithms necessary to sample from default priors,
provided by References \cite{zyczkowski_induced_2001,mezzadri_how_2007,OsiSomZyc10,bruzda_random_2009}.

\begin{figure}[H]
\begin{algorithm}[H]
    \caption{ZS algorithm \cite{zyczkowski_induced_2001} for sampling from the Ginibre matrix ensemble.}
    \label{alg:ginibre-matrix}
    \begin{algorithmic}
        \Require Dimension $D$ and $K$.
        \Ensure An  $D\times K$ matrix, drawn from the Ginibre ensemble.
        \Function{GinibreMatrix}{$D$, $K$}
            \State Generate an $D \times D$ matrix $G$ with elements drawn independently from $(\Nor(0, 1) + \ii \cdot \Nor(0, 1))$.
            \State \Return $G$
        \EndFunction
    \end{algorithmic}
\end{algorithm}
\end{figure}

\begin{figure}[H]
\begin{algorithm}[H]
    \caption{Mezzardi algorithm \cite{mezzadri_how_2007} for sampling from the Haar measure.}
    \label{alg:haar-unitary}
    \begin{algorithmic}
        \Require Dimension $D$.
        \Ensure A unitary operator $U \in \CC^{D\times D}$, drawn from
            the Haar measure.
        \Function{HaarUnitary}{$D$}
            \State $Z \sim \textsc{GinibreMatrix}(D, D)$
            \State $Q, R \gets $ QR-decomposition of $Z$
            \State $\Lambda \gets \diag(R)$
            \State $\lambda_{ii} \gets \lambda_{ii} / |\lambda_{ii}|$
            \State \Return $U \gets Q \Lambda$
        \EndFunction
    \end{algorithmic}
\end{algorithm}
\end{figure}

\begin{figure}[H]
\begin{algorithm}[H]
    \caption{ZS algorithm \cite{zyczkowski_induced_2001} for sampling states from the Ginibre ensemble.}
    \label{alg:ginibre-state}
    \begin{algorithmic}
        \Require Dimension $D$ and rank $K \le D$.
        \Ensure An $D \times D$ positive semidefinite matrix of rank $K$
            drawn from the Ginibre ensemble.
        \Function{GinibreState}{$D$, $K$}
            \State $A \sim \textsc{GinibreMatrix}(D, K)$
            \State $\rho \gets AA^\dagger / \Tr(AA^\dagger)$
            \State \Return $\rho$
        \EndFunction
    \end{algorithmic}
\end{algorithm}
\end{figure}

\begin{figure}[H]
\begin{algorithm}[H]
    \caption{OSZ algorithm \cite{OsiSomZyc10} for sampling from Bures measure.}
    \label{alg:bures-state}
    \begin{algorithmic}
        \Require Dimension $D$.
        \Ensure An $D \times D$ positive semidefinite matrix of rank $K$
            drawn from the Bures measure.
        \Function{BuresState}{$D$}
            \State $A \sim \textsc{GinibreMatrix}(D, D)$
             \State $U\sim \textsc{HaarUnitary}(D)$
             \State $\rho \gets (\Id+ U)AA^\dagger(\Id+ U^\dagger) / \Tr[(\Id+ U)AA^\dagger(\Id+ U^\dagger)]$
            \State \Return $\rho$
        \EndFunction
    \end{algorithmic}
\end{algorithm}
\end{figure}

\begin{figure}[H]
\begin{algorithm}[H]
    \caption{BCSZ algorithm \cite{bruzda_random_2009} for sampling from a uniform ensemble of CPTP maps.}
    \label{alg:random-cptp-map}
    \begin{algorithmic}
        \Require Dimension $D$, Kraus rank $K$
        \Ensure A $D^2 \times D^2$ Choi matrix $J(\Lambda)$ of a channel drawn from the
            BCSZ distribution.
        \Function{BCSZChannel}{$D$}
            \State $X \sim \textsc{GinibreMatrix}(D^2, K)$
            \State $\rho \gets X X^\dagger$
            \State $Y \gets \Tr_2 \rho$
                \Comment{$\Tr_2$ indicates the partial trace over the second copy of $\CC^D$.}
            \State $Z \gets (\id \otimes Y^{-1/2}) \rho (\id \otimes Y^{-1/2})$
            \State \Return $Z D $
        \EndFunction
    \end{algorithmic}
\end{algorithm}
\end{figure}

\section{GADFLI Prior for Coins}
\label{apx:coin-gadfli}

A prior distribution over the bias of a coin is a special case of a qubit and rebit prior. It 
corresponds to a prior over one axis of the Bloch sphere. Without loss of generality, we will take that axis to be the $\hat{z}$-axis,
such that the density operator for a coin is given by
\begin{align}
    \rho = \left(\begin{array}{cc}p & 0 \\0 & 1-p\end{array}\right) =\frac{\Id}{2}+z\frac{Z}{2}= \frac 1 2 \left(\begin{array}{cc}1+z & 0 \\0 & 1-z\end{array}\right),
\end{align}
where $z = 2 p - 1$ is the $\hat{z}$-axis coordinate on the Bloch sphere at which the coin state is positioned.

\begin{figure}[H]
    \begin{center}
        \includegraphics[width=0.6\columnwidth]{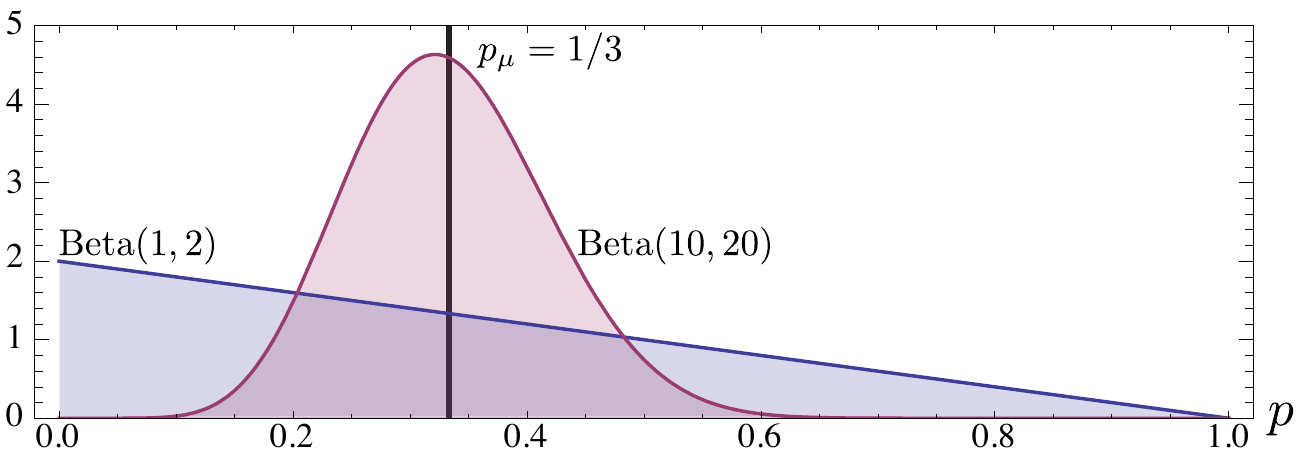}
    \end{center}
    \caption{\label{fig:one-mean-two-priors-coin}
        Two different priors on coins, each with the same prior mean: $p_\mu = 1/3$. This figure should be compared to \fig{two-priors}.
    }
\end{figure}

Given the result of a coin toss $r\in\{0,1\}$, we assign a likelihood for that outcome
conditioned on the coin $p$ as
\begin{align}
    \Pr(r|p)= p^{1-r}(1-p)^{r},
\end{align}
where the probability of heads is $\Pr(0|p)=p$.
Using the density operator definition of a coin, we define that the minimum
eigenvalue of a coin is $\lambda_{\rm min}= \min(p,1-p)$.

\begin{figure}
   \begin{center}
        \includegraphics[width=\columnwidth]{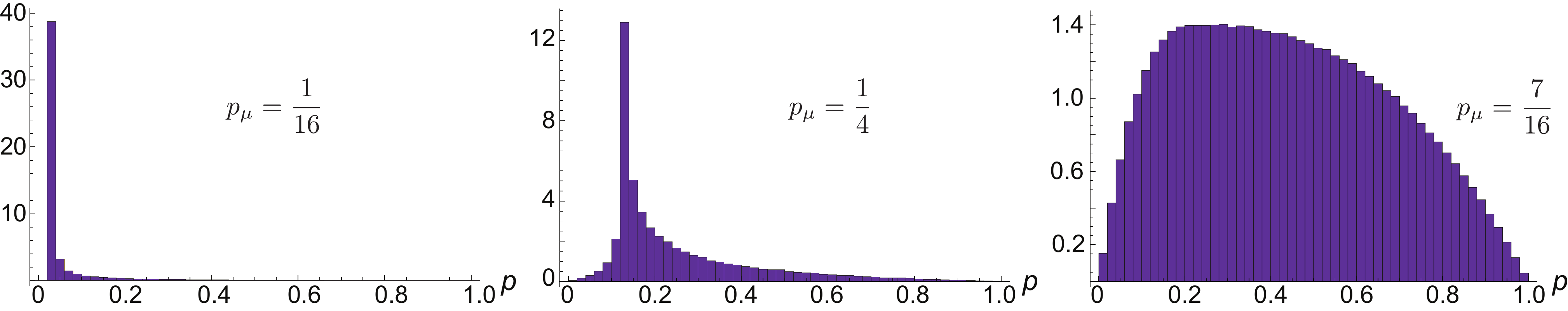}
    \end{center}
    \caption{\label{fig:yay-pretty-priors}
     A histogram of the Generalized Amplitude Damping (GAD) prior for the coin, with 
     $p_\mu \in \{1/16,\,4/16,\,15/16\}$. The histogram was generated with $4,000,000$ 
     samples from the GAD prior. The Gad prior is symmetric about $p_\mu =1/2$ even 
     though $p_\mu =1/2$ is not an allowed value.
    }
\end{figure}

We draw samples of the coin state $p$ from our GAD prior by first choosing 
$\epsilon\sim{\rm Beta}(\alpha,\beta)$ and $p$ to be sampled from a fiducial
prior such that $\expect[p] = 1/2$. Through the rest of this example, we use $p\sim{\rm Uniform}(0,1)$.
The GAD-prior sample $p'$ is obtained by transitioning $p$ with a linear function $\Phi$ given by
\begin{align}
    p' = \Phi[p|\epsilon,p_*] = (1 - \epsilon) p + \epsilon p_*.
\end{align}
The expectation over this GAD prior then gives
the Bayesian mean estimator before any data is collected.
Using that $p$ is fiducial (has mean $1/2$) and that $\expect[\operatorname{Beta}(\alpha, \beta)] = \alpha / (\alpha + \beta)$, we find that
\begin{align}
p_\mu &:= \expect_{p,\epsilon}  [ \Phi(p| \epsilon, p_*)] = \frac{\beta}{\alpha+\beta}\cdot\frac{1}{2} +\frac {  \alpha}{\alpha +\beta} p_*,
\end{align}
and $p_\mu\neq 1/2$. Inverting, we find that the fixed point $p_*$ needed to guarantee that the prior mean is $p_\mu$ is given by
\begin{align}
p_* = \frac {\alpha +\beta}{  \alpha} \left( p_\mu-\frac{\beta}{\alpha+\beta}\cdot\frac{1}{2} \right),
\end{align}
which is analogus to equation \eq{chan_fixed_point}. In order for $p_*$ to be a valid coin, this constrains
\begin{align}
 p_\mu >\frac{\beta}{\alpha+\beta}\cdot\frac{1}{2}.
\end{align}
We find $\alpha$ and $\beta$ consistent with this condition and such that the mean
\begin{align}
\expect_\epsilon[\epsilon]= \frac{\alpha}{\alpha+\beta}
\end{align}
is minimized. This represents that the GAD channel used to define samples does as little as possible
to transform fiducial (uniform) samples to our prior.
Minimizing subject to the constraints that $\alpha>0,\beta>0$ and  $p_*>0 $, we obtain that
\begin{align}
\alpha = 1 \quad {\rm and}\quad \beta
 =  \begin{dcases}
\frac{1}{2 p_\mu -1}-1 & \text{if } p_\mu>1/2\\ \vspace{5pt}
\frac{2 p_\mu}{1-2 p_\mu } & \text{if }p_\mu<1/2
\end{dcases}.
\end{align} 
We can write this in terms of $\lambda_{\min}$ to obtain the final GAD-prior condition,
\begin{align}
\alpha = 1 \quad {\rm and}\quad \beta
 = \frac{2 \lambda_{\min}}{1-2 \lambda_{\min} },
\end{align}
which should be compared to \eq{gadfli_beta}.
Using this condition, we obtain the coin priors shown in \fig{yay-pretty-priors}.

\section{An estimate of the tracking frequency}
\label{apx:track_freq}

We now give an order of magnitude estimate for maximum frequency that is
``smoothly'' trackable by our algorithm. It is possible our algorithm can
track higher frequencies but the evolution is likely to be fairly
discontinuous.

Our estimate for the tracking frequency comes from sampling arguments in
tracking a bias of a coin, with probability $\Pr(0|p)=p$ for heads. We will
assume the coin's bias changes from $\Pr(0|p)=p$ to $\Pr(0|p)=p \pm\epsilon$
where $\epsilon\ll 1$. The question is how quickly can we detect such a
change, i.e. how many equally spaced in time samples does it take to notice
the difference of $\pm \epsilon$.

We must specify an error tolerance--$\epsilon$--for our sensing protocol and a
confidence interval--$Z$--for our estimate of $p$. It turns out if we use the
trace distance between two coins $\rho = \diag(p,1-p)$ and $\sigma =\diag(p\pm
\epsilon, 1-p\pm \epsilon)$ then $D(\rho,\sigma)= (1/2) \Tr{|\rho -\sigma|}=
\epsilon$. These two specifications will in turn (approximately) determine the
number of samples required and therefore the bandwidth of our tracking
protocol. We choose $Z=1.9599$ which corresponds to a $95\%$ level of
confidence for our estimator and the maximum acceptable error $|p_{\rm true} -
p_{\rm est}|<\epsilon$. Using the standard deviation of the Bernouli
distribution $\sqrt{p(1-p)/N}$, we have $\epsilon = Z \sqrt{p(1-p)/N}$. The
largest standard deviation (worst case) is when $p=1/2$ this gives $\epsilon =
Z /(2\sqrt{N})$. Rearranging for $N$ gives $N=Z^2 /(4\epsilon^2)$. To obtain
this many samples we must measure for a time $T_{\rm meas}= \Delta t N$, where
$\Delta t $ is the time between measurements. Thus $T_{\rm meas}$ is
effectively the time between our samples of $p(t)$. Na\"ive arguments from the
Nyquist-Shannon sampling theorem imply that we can not determine frequency
components of $p(t)$ greater than $f_{\rm max}=1/(2T_{\rm meas})$, which is
called the is detection frequency bandwidth. The implication is we can track,
for example, $p(t)= 1/2 +\epsilon \sin(f_{\rm max})$. 

\section{QInfer Tomography Tutorial}
\label{apx:qinfer-tutorial}

    In this Appendix, we demonstrate the use of
\href{https://github.com/QInfer/python-qinfer}{QInfer} for Bayesian
state and process tomography. In particular, we show how to estimate
states and channels given data synthesized from a description of a true
state, and discuss how to obtain region estimates, covariance
superoperators and other useful functions of tomography posteriors. We
then discuss how to apply these techniques in experimental systems.

    The tomography implementation in QInfer is based on
\href{http://qutip.org/}{QuTiP} and \href{numpy.org}{NumPy}, so we start
by importing everything here.

    \begin{Verbatim}[commandchars=\\\{\}]
{\color{incolor}In [{\color{incolor}1}]:} \PY{k+kn}{import} \PY{n+nn}{numpy} \PY{k+kn}{as} \PY{n+nn}{np}
        \PY{k+kn}{import} \PY{n+nn}{qutip} \PY{k+kn}{as} \PY{n+nn}{qt}
        \PY{k+kn}{import} \PY{n+nn}{qinfer} \PY{k+kn}{as} \PY{n+nn}{qi}
\end{Verbatim}

    As a first step, we define a \emph{basis} for performing tomography; the
choice of basis is largely arbitrary, but depending on the experiment,
some bases may be more or less convienent. Here, we focus on the example
of the single-qubit Pauli basis
\(B = \{\id, \sigma_x, \sigma_y, \sigma_z\}\).

    \begin{Verbatim}[commandchars=\\\{\}]
{\color{incolor}In [{\color{incolor}2}]:} \PY{n}{basis} \PY{o}{=} \PY{n}{qi}\PY{o}{.}\PY{n}{tomography}\PY{o}{.}\PY{n}{pauli\PYZus{}basis}\PY{p}{(}\PY{l+m+mi}{1}\PY{p}{)}
        \PY{n}{display}\PY{p}{(}\PY{n}{basis}\PY{p}{)}
\end{Verbatim}

    \begin{verbatim}
<TomographyBasis dims=[2] at 362504488>
    \end{verbatim}

    We will get a lot of use out of the Pauli basis, so we also define some
useful shorthand.

    \begin{Verbatim}[commandchars=\\\{\}]
{\color{incolor}In [{\color{incolor}3}]:} \PY{n}{I}\PY{p}{,} \PY{n}{X}\PY{p}{,} \PY{n}{Y}\PY{p}{,} \PY{n}{Z} \PY{o}{=} \PY{n}{qt}\PY{o}{.}\PY{n}{qeye}\PY{p}{(}\PY{l+m+mi}{2}\PY{p}{)}\PY{p}{,} \PY{n}{qt}\PY{o}{.}\PY{n}{sigmax}\PY{p}{(}\PY{p}{)}\PY{p}{,} \PY{n}{qt}\PY{o}{.}\PY{n}{sigmay}\PY{p}{(}\PY{p}{)}\PY{p}{,} \PY{n}{qt}\PY{o}{.}\PY{n}{sigmaz}\PY{p}{(}\PY{p}{)}
\end{Verbatim}

    Basis objects are responsible for converting between QuTiP's rich
\texttt{Qobj} format and the unstructured model parameter representation
used by QInfer.

    \begin{Verbatim}[commandchars=\\\{\}]
{\color{incolor}In [{\color{incolor}4}]:} \PY{n}{display}\PY{p}{(}\PY{n}{basis}\PY{o}{.}\PY{n}{state\PYZus{}to\PYZus{}modelparams}\PY{p}{(}\PY{n}{I} \PY{o}{/} \PY{l+m+mi}{2} \PY{o}{+} \PY{n}{X} \PY{o}{/} \PY{l+m+mi}{2}\PY{p}{)}\PY{p}{)}
\end{Verbatim}

    \begin{verbatim}
array([ 0.70710678,  0.70710678,  0.        ,  0.        ])
    \end{verbatim}

    \begin{Verbatim}[commandchars=\\\{\}]
{\color{incolor}In [{\color{incolor}5}]:} \PY{n}{display}\PY{p}{(}\PY{n}{basis}\PY{o}{.}\PY{n}{modelparams\PYZus{}to\PYZus{}state}\PY{p}{(}\PY{n}{np}\PY{o}{.}\PY{n}{array}\PY{p}{(}\PY{p}{[}\PY{l+m+mi}{1}\PY{p}{,} \PY{l+m+mi}{0}\PY{p}{,} \PY{l+m+mi}{0}\PY{p}{,} \PY{l+m+mi}{1}\PY{p}{]}\PY{p}{)} \PY{o}{/} \PY{n}{np}\PY{o}{.}\PY{n}{sqrt}\PY{p}{(}\PY{l+m+mi}{2}\PY{p}{)}\PY{p}{)}\PY{p}{)}
\end{Verbatim}

    Quantum object: dims = [[2], [2]], shape = [2, 2], type = oper, isherm = True\begin{equation*}\left(\begin{array}{*{11}c}1.000 & 0.0\\0.0 & 0.0\\\end{array}\right)\end{equation*}

    Having defined a basis, we then define the core object describing a
tomography experiment, the \emph{model}. In QInfer, models encapsulate
the likelihood function, experimental parameters and other useful
metadata about the experimental properties being estimated. In our case,
we use \texttt{TomographyModel} to describe the single-shot experiment,
and \texttt{BinomialModel} to describe batches of the single-shot
experiment.

    \begin{Verbatim}[commandchars=\\\{\}]
{\color{incolor}In [{\color{incolor}6}]:} \PY{n}{model} \PY{o}{=} \PY{n}{qi}\PY{o}{.}\PY{n}{BinomialModel}\PY{p}{(}\PY{n}{qi}\PY{o}{.}\PY{n}{tomography}\PY{o}{.}\PY{n}{TomographyModel}\PY{p}{(}\PY{n}{basis}\PY{p}{)}\PY{p}{)}
         \PY{n}{display}\PY{p}{(}\PY{n}{model}\PY{p}{)}
\end{Verbatim}

    \begin{verbatim}
<qinfer.derived_models.BinomialModel at 0x159b6048>
    \end{verbatim}

    A \texttt{Model} defines a vector of model parameters; for a single
qubit \texttt{TomographyModel}, this is a vector of length 4, each
describing a different element of the Hermitian operator basis. Each
\texttt{Model} also defines experiment parameters as a NumPy record
array. A record then describes a single measurement of the model.

    \begin{Verbatim}[commandchars=\\\{\}]
{\color{incolor}In [{\color{incolor}7}]:} \PY{n}{display}\PY{p}{(}\PY{n}{model}\PY{o}{.}\PY{n}{expparams\PYZus{}dtype}\PY{p}{)}
\end{Verbatim}

    \begin{verbatim}
[('meas', float, 4), ('n_meas', 'uint')]
    \end{verbatim}

    In this case, the experiment parameters record has two \emph{fields}:
\texttt{meas} and \texttt{n\_meas}. The first is a vector of four floats
corresponding to
\(\sket{M} = (\sbraket{B_0 | M}, \sbraket{B_1 | M}, \sbraket{B_2 | M}, \sbraket{B_3 | M)}\).
The second is an unsigned integer (\texttt{uint}) describing how many
times that measurement is performed. For instance, measuring
\((\id + \sigma_z) / 2\) 40 times is given by the array:

    \begin{Verbatim}[commandchars=\\\{\}]
{\color{incolor}In [{\color{incolor}8}]:} \PY{n}{expparams} \PY{o}{=} \PY{n}{np}\PY{o}{.}\PY{n}{array}\PY{p}{(}\PY{p}{[}
             \PY{c}{\PYZsh{} Each tuple, marked with (), defines a single record.}
             \PY{p}{(}
                 \PY{c}{\PYZsh{} Within each tuple, fields are separated by commas.}
                 \PY{c}{\PYZsh{} The fields follow in the order given by the model,}
                 \PY{c}{\PYZsh{} so the first field is meas, a length\PYZhy{}4 vector.}
                 \PY{p}{[}\PY{l+m+mi}{1} \PY{o}{/} \PY{n}{np}\PY{o}{.}\PY{n}{sqrt}\PY{p}{(}\PY{l+m+mi}{2}\PY{p}{)}\PY{p}{,} \PY{l+m+mi}{0}\PY{p}{,} \PY{l+m+mi}{0}\PY{p}{,} \PY{l+m+mi}{1} \PY{o}{/} \PY{n}{np}\PY{o}{.}\PY{n}{sqrt}\PY{p}{(}\PY{l+m+mi}{2}\PY{p}{)}\PY{p}{]}\PY{p}{,}
                 \PY{c}{\PYZsh{} The second field is then the number of measurements.}
                 \PY{l+m+mi}{40}
             \PY{p}{)}
         \PY{p}{]}\PY{p}{,}
         \PY{c}{\PYZsh{} We finish building the array by passing along the right data\PYZbs{}}
         \PY{c}{\PYZsh{} type to NumPy. This is somwhat of a QInfer idiom.}
         \PY{n}{dtype}\PY{o}{=}\PY{n}{model}\PY{o}{.}\PY{n}{expparams\PYZus{}dtype}\PY{p}{)}
         \PY{n}{display}\PY{p}{(}\PY{n}{expparams}\PY{p}{)}
\end{Verbatim}

    \begin{verbatim}
array([([0.7071067811865475, 0.0, 0.0, 0.7071067811865475], 40L)], 
      dtype=[('meas', '<f8', (4,)), ('n_meas', '<u4')])
    \end{verbatim}

    The fields of a record array can be obtained by indexing. For instance,
the \texttt{{[}\textquotesingle{}meas\textquotesingle{}{]}} field is
then a \(1 \times 4\) array, with the first index allowing for a
sequence of measurements to be described at once.

    \begin{Verbatim}[commandchars=\\\{\}]
{\color{incolor}In [{\color{incolor}9}]:} \PY{n}{display}\PY{p}{(}\PY{n}{expparams}\PY{p}{[}\PY{l+s}{\PYZsq{}}\PY{l+s}{meas}\PY{l+s}{\PYZsq{}}\PY{p}{]}\PY{p}{)}
\end{Verbatim}

    \begin{verbatim}
array([[ 0.70710678,  0.        ,  0.        ,  0.70710678]])
    \end{verbatim}

    Note that by convention, \texttt{meas} is normalized to
\(1 / \sqrt{d}\).

    Often, we will not construct experiments directly, but will instead rely
on QInfer's heuristics (described below). In any case, once we have a
model, the next step is to create a prior. QInfer comes with several
useful fiducial priors, as well as insightful priors constructed from
amplitude damping channels. For instance, to create a Hilbert-Schmidt
uniform prior constrained to rebits, we use the
\texttt{GinibreReditDistribution}:

    \begin{Verbatim}[commandchars=\\\{\}]
{\color{incolor}In [{\color{incolor}10}]:} \PY{n}{fiducial\PYZus{}prior} \PY{o}{=} \PY{n}{qi}\PY{o}{.}\PY{n}{tomography}\PY{o}{.}\PY{n}{GinibreReditDistribution}\PY{p}{(}\PY{n}{basis}\PY{p}{)}
\end{Verbatim}

    \begin{Verbatim}[commandchars=\\\{\}]
{\color{incolor}In [{\color{incolor}11}]:} \PY{n}{qi}\PY{o}{.}\PY{n}{tomography}\PY{o}{.}\PY{n}{plotting\PYZus{}tools}\PY{o}{.}\PY{n}{plot\PYZus{}rebit\PYZus{}prior}\PY{p}{(}\PY{n}{fiducial\PYZus{}prior}\PY{p}{,} \PY{n}{rebit\PYZus{}axes}\PY{o}{=}\PY{p}{[}\PY{l+m+mi}{1}\PY{p}{,} \PY{l+m+mi}{3}\PY{p}{]}\PY{p}{)}
\end{Verbatim}

    \begin{center}
    \adjustimage{max size={0.6\linewidth}{}}{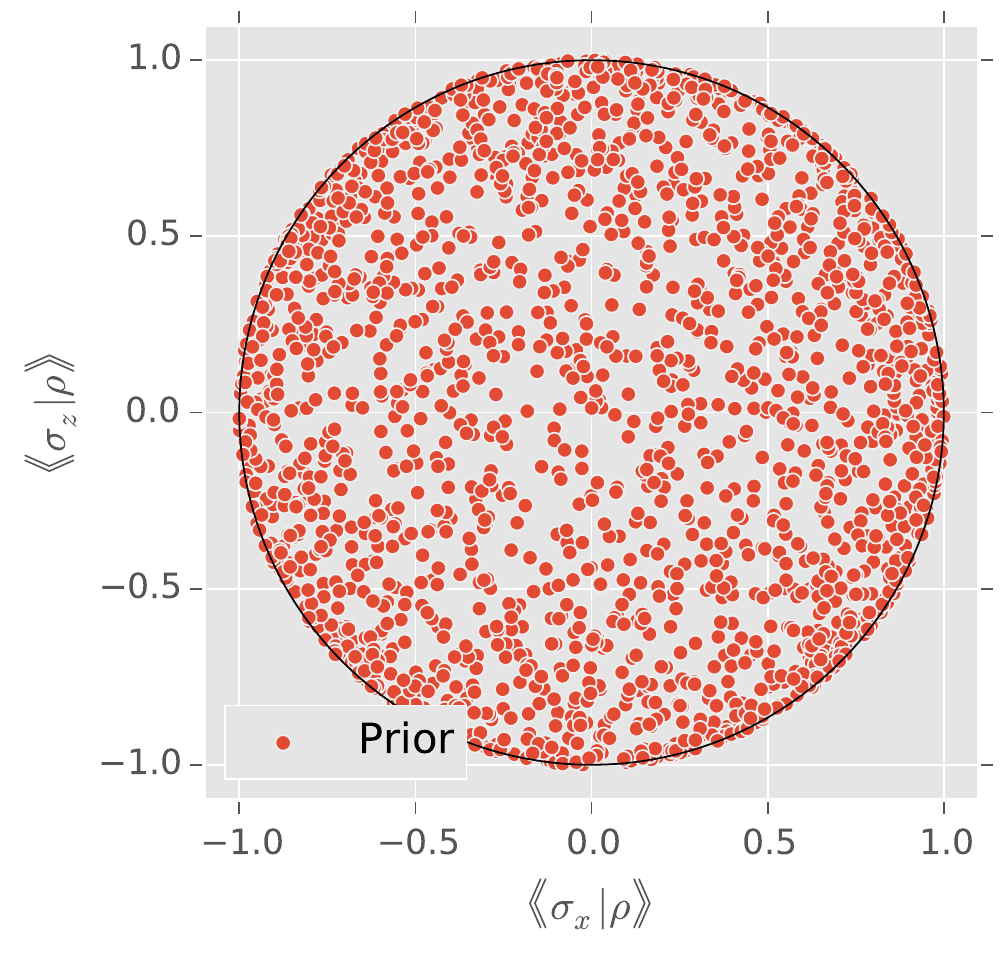}
    \end{center}
    { \hspace*{\fill} \\}
    
    Here, we have told QInfer that we wish to treat \(\sigma_x\) and
\(\sigma_z\) as our rebit axes using the
\texttt{rebit\_axes={[}1,\ 3{]}} argument.

    Insightful priors can be constructed by specifying a fiducial
prior and a QuTiP \texttt{Qobj} representing the desired
mean.

    \begin{Verbatim}[commandchars=\\\{\}]
{\color{incolor}In [{\color{incolor}12}]:} \PY{n}{prior\PYZus{}mean} \PY{o}{=} \PY{p}{(}\PY{n}{I} \PY{o}{+} \PY{p}{(}\PY{l+m+mi}{2}\PY{o}{/}\PY{l+m+mi}{3}\PY{p}{)} \PY{o}{*} \PY{n}{Z} \PY{o}{+} \PY{p}{(}\PY{l+m+mi}{1}\PY{o}{/}\PY{l+m+mi}{3}\PY{p}{)} \PY{o}{*} \PY{n}{X}\PY{p}{)} \PY{o}{/} \PY{l+m+mi}{2}
         \PY{n}{display}\PY{p}{(}\PY{n}{prior\PYZus{}mean}\PY{p}{)}
\end{Verbatim}

    Quantum object: dims = [[2], [2]], shape = [2, 2], type = oper, isherm = True\begin{equation*}\left(\begin{array}{*{11}c}0.833 & 0.167\\0.167 & 0.167\\\end{array}\right)\end{equation*}

    \begin{Verbatim}[commandchars=\\\{\}]
{\color{incolor}In [{\color{incolor}13}]:} \PY{n}{prior} \PY{o}{=} \PY{n}{qi}\PY{o}{.}\PY{n}{tomography}\PY{o}{.}\PY{n}{GADFLIDistribution}\PY{p}{(}\PY{n}{fiducial\PYZus{}prior}\PY{p}{,} \PY{n}{prior\PYZus{}mean}\PY{p}{)}
\end{Verbatim}

    \begin{Verbatim}[commandchars=\\\{\}]
{\color{incolor}In [{\color{incolor}14}]:} \PY{n}{qi}\PY{o}{.}\PY{n}{tomography}\PY{o}{.}\PY{n}{plotting\PYZus{}tools}\PY{o}{.}\PY{n}{plot\PYZus{}rebit\PYZus{}prior}\PY{p}{(}\PY{n}{prior}\PY{p}{,} \PY{n}{rebit\PYZus{}axes}\PY{o}{=}\PY{p}{[}\PY{l+m+mi}{1}\PY{p}{,} \PY{l+m+mi}{3}\PY{p}{]}\PY{p}{)}
\end{Verbatim}

    \begin{center}
    \adjustimage{max size={0.6\linewidth}{}}{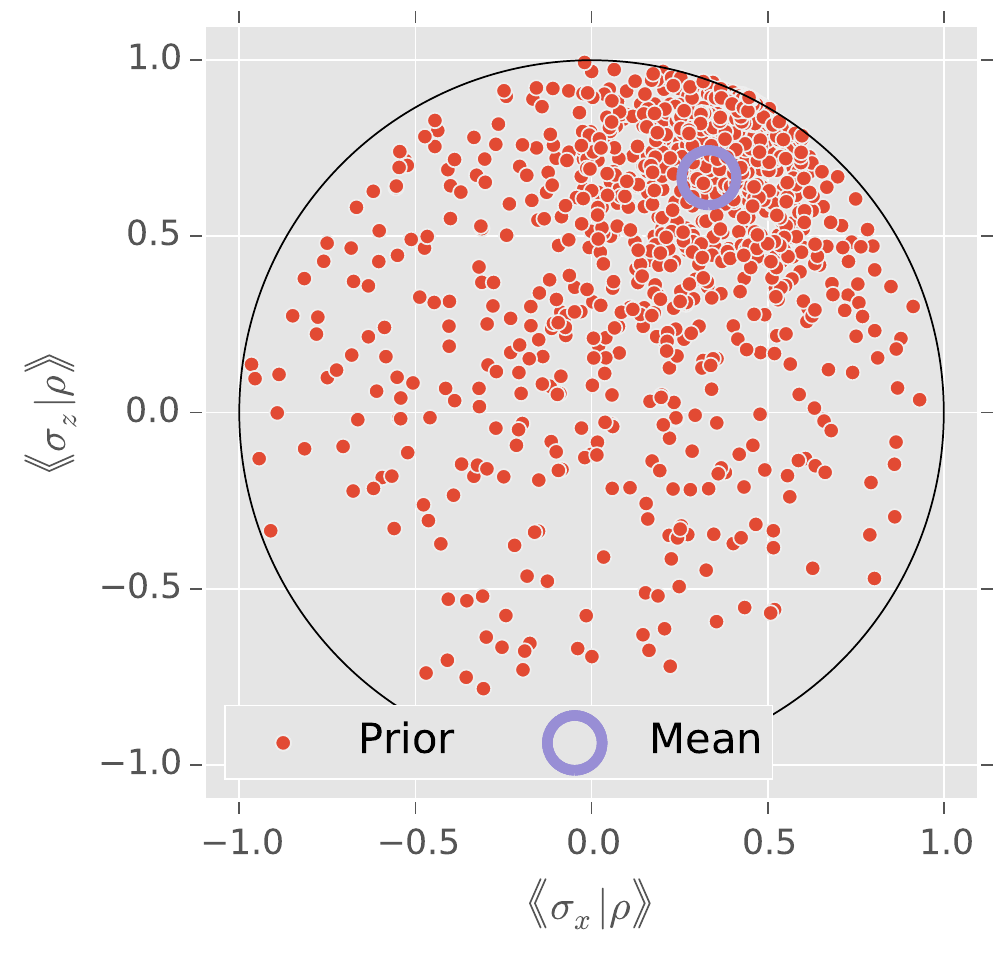}
    \end{center}
    { \hspace*{\fill} \\}
    
    Having constructed a prior and a model, we can now continue to perform
Bayesian inference using SMC. We demonstrate using the true state
\(\rho = \id / 2 + (2 / 3) \sigma_z / 2\) with the prior mean
\(\rho_\mu = \id / 2 + (4/5) \sigma_z + (1/7)\sigma_x\).

    \begin{Verbatim}[commandchars=\\\{\}]
{\color{incolor}In [{\color{incolor}15}]:} \PY{n}{basis} \PY{o}{=} \PY{n}{qi}\PY{o}{.}\PY{n}{tomography}\PY{o}{.}\PY{n}{pauli\PYZus{}basis}\PY{p}{(}\PY{l+m+mi}{1}\PY{p}{)}
         \PY{n}{model} \PY{o}{=} \PY{n}{qi}\PY{o}{.}\PY{n}{BinomialModel}\PY{p}{(}\PY{n}{qi}\PY{o}{.}\PY{n}{tomography}\PY{o}{.}\PY{n}{TomographyModel}\PY{p}{(}\PY{n}{basis}\PY{p}{)}\PY{p}{)}
         \PY{n}{true\PYZus{}state} \PY{o}{=} \PY{n}{basis}\PY{o}{.}\PY{n}{state\PYZus{}to\PYZus{}modelparams}\PY{p}{(}
             \PY{n}{I} \PY{o}{/} \PY{l+m+mi}{2} \PY{o}{+} \PY{p}{(}\PY{l+m+mi}{2} \PY{o}{/} \PY{l+m+mi}{3}\PY{p}{)} \PY{o}{*} \PY{n}{Z} \PY{o}{/} \PY{l+m+mi}{2}
         \PY{p}{)}\PY{p}{[}\PY{n}{np}\PY{o}{.}\PY{n}{newaxis}\PY{p}{,} \PY{p}{:}\PY{p}{]}
         \PY{n}{fiducial\PYZus{}prior} \PY{o}{=} \PY{n}{qi}\PY{o}{.}\PY{n}{tomography}\PY{o}{.}\PY{n}{GinibreReditDistribution}\PY{p}{(}\PY{n}{basis}\PY{p}{)}
         \PY{n}{prior} \PY{o}{=} \PY{n}{qi}\PY{o}{.}\PY{n}{tomography}\PY{o}{.}\PY{n}{GADFLIDistribution}\PY{p}{(}\PY{n}{fiducial\PYZus{}prior}\PY{p}{,}
             \PY{n}{I} \PY{o}{/} \PY{l+m+mi}{2} \PY{o}{+} \PY{p}{(}\PY{l+m+mi}{4} \PY{o}{/} \PY{l+m+mi}{5}\PY{p}{)} \PY{o}{*} \PY{n}{Z} \PY{o}{/} \PY{l+m+mi}{2} \PY{o}{+} \PY{p}{(}\PY{l+m+mi}{1} \PY{o}{/} \PY{l+m+mi}{7}\PY{p}{)} \PY{o}{*} \PY{n}{X} \PY{o}{/} \PY{l+m+mi}{2}
         \PY{p}{)}
\end{Verbatim}

    \begin{Verbatim}[commandchars=\\\{\}]
{\color{incolor}In [{\color{incolor}16}]:} \PY{n}{qi}\PY{o}{.}\PY{n}{tomography}\PY{o}{.}\PY{n}{plotting\PYZus{}tools}\PY{o}{.}\PY{n}{plot\PYZus{}rebit\PYZus{}prior}\PY{p}{(}\PY{n}{prior}\PY{p}{,} \PY{n}{true\PYZus{}state}\PY{o}{=}\PY{n}{true\PYZus{}state}\PY{p}{,} \PY{n}{rebit\PYZus{}axes}\PY{o}{=}\PY{p}{[}\PY{l+m+mi}{1}\PY{p}{,} \PY{l+m+mi}{3}\PY{p}{]}\PY{p}{)}
\end{Verbatim}

    \begin{center}
    \adjustimage{max size={0.6\linewidth}{}}{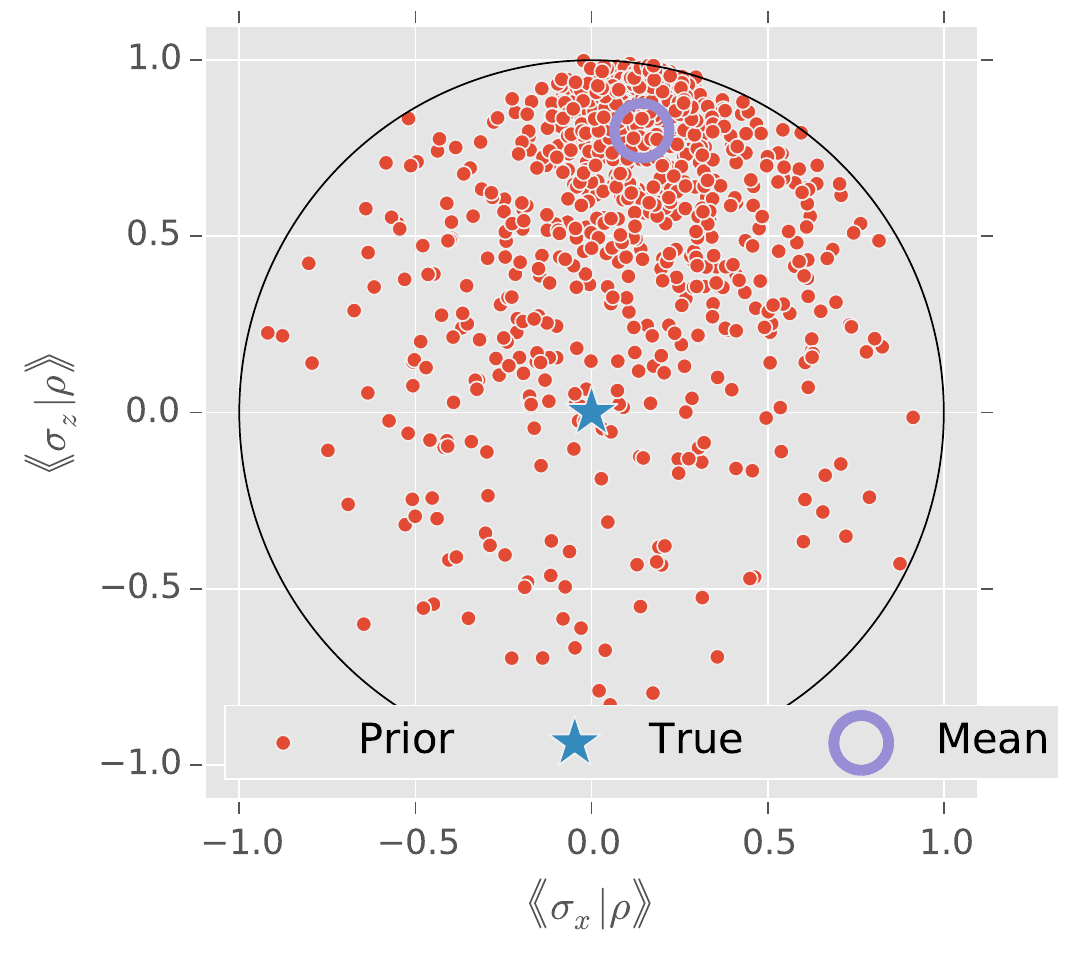}
    \end{center}
    { \hspace*{\fill} \\}
    
    The updater and heuristic classes track the posterior and the
random-measurement experiment design, respectively.

    \begin{Verbatim}[commandchars=\\\{\}]
{\color{incolor}In [{\color{incolor}17}]:} \PY{n}{updater} \PY{o}{=} \PY{n}{qi}\PY{o}{.}\PY{n}{smc}\PY{o}{.}\PY{n}{SMCUpdater}\PY{p}{(}\PY{n}{model}\PY{p}{,} \PY{l+m+mi}{2000}\PY{p}{,} \PY{n}{prior}\PY{p}{)}
         \PY{n}{heuristic} \PY{o}{=} \PY{n}{qi}\PY{o}{.}\PY{n}{tomography}\PY{o}{.}\PY{n}{RandomPauliHeuristic}\PY{p}{(}\PY{n}{updater}\PY{p}{,} \PY{n}{other\PYZus{}fields}\PY{o}{=}\PY{p}{\PYZob{}}\PY{l+s}{\PYZsq{}}\PY{l+s}{n\PYZus{}meas}\PY{l+s}{\PYZsq{}}\PY{p}{:} \PY{l+m+mi}{40}\PY{p}{\PYZcb{}}\PY{p}{)}
\end{Verbatim}

    We synthesize data for the true state, then feed it into the updater in
order to obtain our final posterior.

    \begin{Verbatim}[commandchars=\\\{\}]
{\color{incolor}In [{\color{incolor}18}]:} \PY{k}{for} \PY{n}{idx\PYZus{}exp} \PY{o+ow}{in} \PY{n+nb}{xrange}\PY{p}{(}\PY{l+m+mi}{50}\PY{p}{)}\PY{p}{:}
             \PY{n}{experiment} \PY{o}{=} \PY{n}{heuristic}\PY{p}{(}\PY{p}{)}
             \PY{n}{datum} \PY{o}{=} \PY{n}{model}\PY{o}{.}\PY{n}{simulate\PYZus{}experiment}\PY{p}{(}\PY{n}{true\PYZus{}state}\PY{p}{,} \PY{n}{experiment}\PY{p}{)}
             \PY{n}{updater}\PY{o}{.}\PY{n}{update}\PY{p}{(}\PY{n}{datum}\PY{p}{,} \PY{n}{experiment}\PY{p}{)}
\end{Verbatim}

    \begin{Verbatim}[commandchars=\\\{\}]
{\color{incolor}In [{\color{incolor}19}]:} \PY{n}{plt}\PY{o}{.}\PY{n}{figure}\PY{p}{(}\PY{n}{figsize}\PY{o}{=}\PY{p}{(}\PY{l+m+mi}{10}\PY{p}{,} \PY{l+m+mi}{10}\PY{p}{)}\PY{p}{)}
         \PY{n}{qi}\PY{o}{.}\PY{n}{tomography}\PY{o}{.}\PY{n}{plotting\PYZus{}tools}\PY{o}{.}\PY{n}{plot\PYZus{}rebit\PYZus{}posterior}\PY{p}{(}
             \PY{n}{updater}\PY{p}{,} \PY{n}{prior}\PY{p}{,} \PY{n}{true\PYZus{}state}\PY{p}{,}
             \PY{n}{rebit\PYZus{}axes}\PY{o}{=}\PY{p}{[}\PY{l+m+mi}{1}\PY{p}{,} \PY{l+m+mi}{3}\PY{p}{]}
         \PY{p}{)}
\end{Verbatim}

    \begin{center}
    \adjustimage{max size={0.6\linewidth}{}}{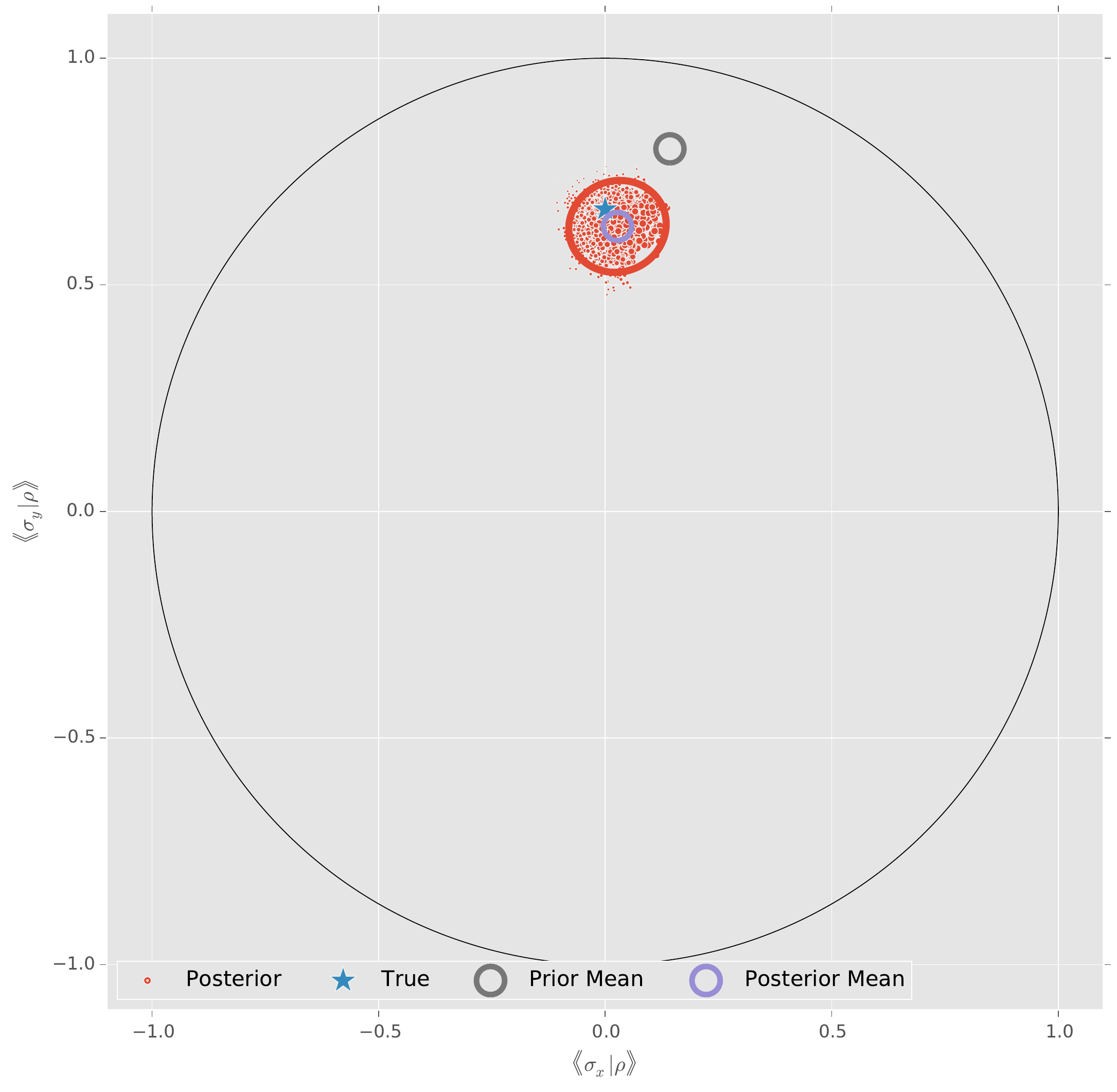}
    \end{center}
    { \hspace*{\fill} \\}
    
    We can use our tomography basis object to read out the estimated final
state as a QuTiP \texttt{Qobj}.

    \begin{Verbatim}[commandchars=\\\{\}]
{\color{incolor}In [{\color{incolor}20}]:} \PY{n}{est\PYZus{}mean} \PY{o}{=} \PY{n}{basis}\PY{o}{.}\PY{n}{modelparams\PYZus{}to\PYZus{}state}\PY{p}{(}\PY{n}{updater}\PY{o}{.}\PY{n}{est\PYZus{}mean}\PY{p}{(}\PY{p}{)}\PY{p}{)}
         \PY{n}{display}\PY{p}{(}\PY{n}{est\PYZus{}mean}\PY{p}{)}
\end{Verbatim}

    Quantum object: dims = [[2], [2]], shape = [2, 2], type = oper, isherm = True\begin{equation*}\left(\begin{array}{*{11}c}0.814 & 0.014\\0.014 & 0.186\\\end{array}\right)\end{equation*}

    As discussed in the main text, the posterior can also be described by
the covariance superoperator \(\Sigma \rho = \Cov(\sket{\rho})\). We
demonstrate by showing the Choi matrix \(J(\Cov(\sket{\rho}))\).

    \begin{Verbatim}[commandchars=\\\{\}]
{\color{incolor}In [{\color{incolor}21}]:} \PY{n}{cov\PYZus{}superop} \PY{o}{=} \PY{n}{basis}\PY{o}{.}\PY{n}{covariance\PYZus{}mtx\PYZus{}to\PYZus{}superop}\PY{p}{(}\PY{n}{updater}\PY{o}{.}\PY{n}{est\PYZus{}covariance\PYZus{}mtx}\PY{p}{(}\PY{p}{)}\PY{p}{)}
         \PY{n}{display}\PY{p}{(}\PY{n}{qt}\PY{o}{.}\PY{n}{to\PYZus{}choi}\PY{p}{(}\PY{n}{cov\PYZus{}superop}\PY{p}{)}\PY{p}{)}
         \PY{n}{display}\PY{p}{(}\PY{n}{Latex}\PY{p}{(}\PY{l+s}{r\PYZdq{}}\PY{l+s}{\PYZdl{}}\PY{l+s}{\PYZbs{}}\PY{l+s}{|}\PY{l+s}{\PYZbs{}}\PY{l+s}{Sigma}\PY{l+s}{\PYZbs{}}\PY{l+s}{rho}\PY{l+s}{\PYZbs{}}\PY{l+s}{|\PYZus{}\PYZob{}\PYZob{}}\PY{l+s}{\PYZbs{}}\PY{l+s}{Tr\PYZcb{}\PYZcb{} = \PYZob{}:0.4f\PYZcb{}\PYZdl{}}\PY{l+s}{\PYZdq{}}\PY{o}{.}\PY{n}{format}\PY{p}{(}\PY{n}{cov\PYZus{}superop}\PY{o}{.}\PY{n}{norm}\PY{p}{(}\PY{l+s}{\PYZsq{}}\PY{l+s}{tr}\PY{l+s}{\PYZsq{}}\PY{p}{)}\PY{p}{)}\PY{p}{)}\PY{p}{)}
\end{Verbatim}

    Quantum object: dims = [[[2], [2]], [[2], [2]]], shape = [4, 4], type = super, isherm = True, superrep = choi\begin{equation*}\left(\begin{array}{*{11}c}2.854\times10^{-04} & 1.622\times10^{-05} & 1.622\times10^{-05} & 3.210\times10^{-04}\\1.622\times10^{-05} & -2.854\times10^{-04} & 3.210\times10^{-04} & -1.622\times10^{-05}\\1.622\times10^{-05} & 3.210\times10^{-04} & -2.854\times10^{-04} & -1.622\times10^{-05}\\3.210\times10^{-04} & -1.622\times10^{-05} & -1.622\times10^{-05} & 2.854\times10^{-04}\\\end{array}\right)\end{equation*}

        \begin{equation*}\adjustbox{max width=\hsize}{$
        \|\Sigma\rho\|_{\Tr} = 0.0012
        $}\end{equation*}

    Here, we use the Hinton diagram plotting functionality provided by QuTiP
to depict the covariance in each observable that we obtain from the
posterior.

    \begin{Verbatim}[commandchars=\\\{\}]
{\color{incolor}In [{\color{incolor}22}]:} \PY{n}{display}\PY{p}{(}\PY{n}{qt}\PY{o}{.}\PY{n}{visualization}\PY{o}{.}\PY{n}{hinton}\PY{p}{(}\PY{n}{cov\PYZus{}superop}\PY{p}{)}\PY{p}{)}
\end{Verbatim}

    \begin{verbatim}
(<matplotlib.figure.Figure at 0x18218a20>,
 <matplotlib.axes._subplots.AxesSubplot at 0x18209f98>)
    \end{verbatim}

    \begin{center}
    \adjustimage{max size={0.6\linewidth}{}}{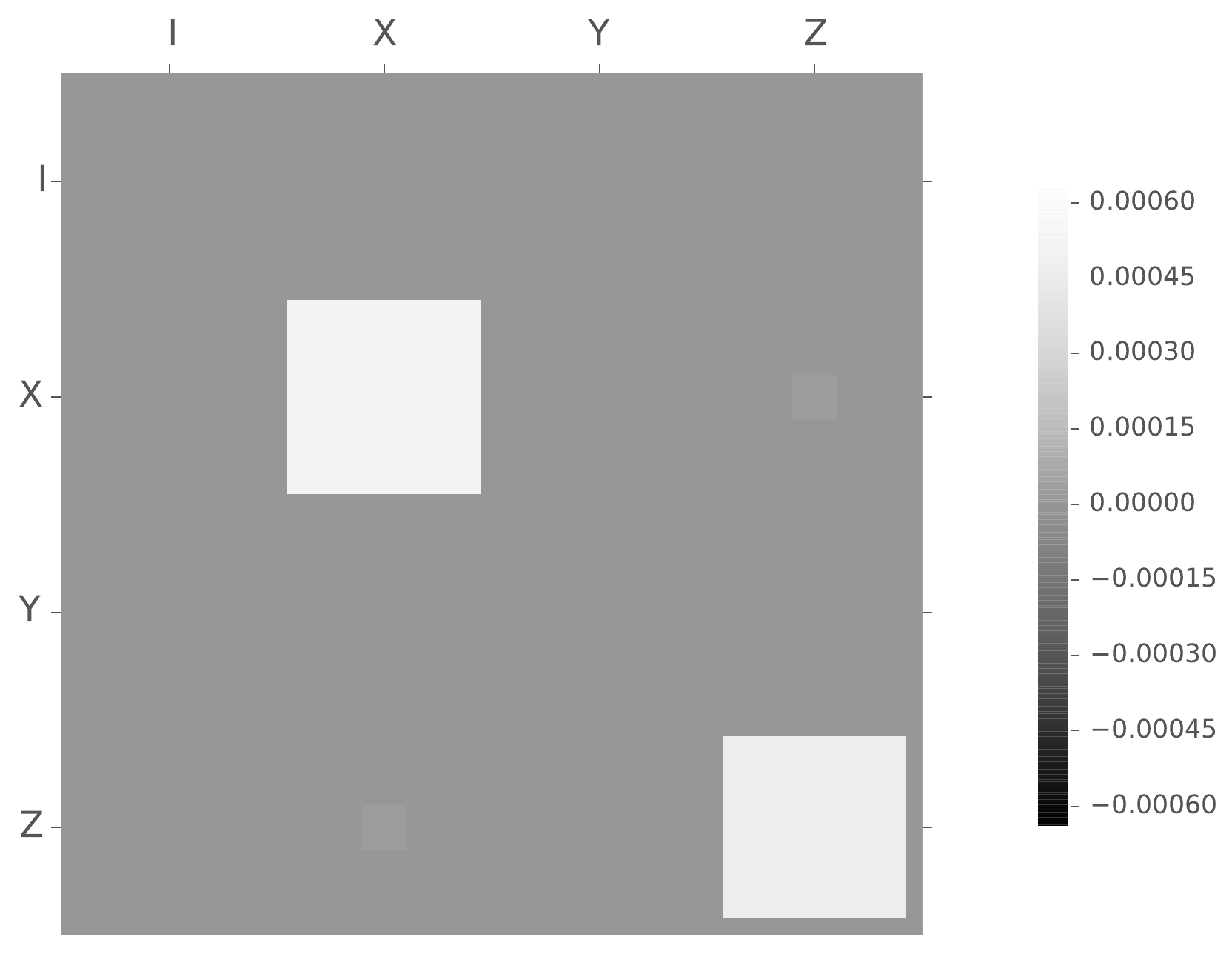}
    \end{center}
    { \hspace*{\fill} \\}

\end{document}